\documentclass[11pt]{article}
\usepackage{epsfig}
\usepackage{amssymb}
\newcommand{\sect}[1]{ \section{#1} \setcounter{equation}{0} }

\newcommand{\Dslash}{D \! \! \! \! /}

\newcommand{\half}{\mbox{\small{$\frac{1}{2}$}}}

\newcommand{\MSbar}{\overline{\mbox{MS}}}

\newcommand{\Nc}{N_{\!c}}
\newcommand{\Nf}{N_{\!f}}
\newcommand{\NF}{N_{\!F}}
\newcommand{\NA}{N_{\!A}}

\begin{document}
\title{Eight dimensional QCD at one loop}
\author{J.A. Gracey, \\ Theoretical Physics Division, \\ 
Department of Mathematical Sciences, \\ University of Liverpool, \\ P.O. Box 
147, \\ Liverpool, \\ L69 3BX, \\ United Kingdom.} 
\date{}
\maketitle 

\vspace{5cm} 
\noindent 
{\bf Abstract.} The Lagrangian for a non-abelian gauge theory with an $SU(\Nc)$
symmetry and a linear covariant gauge fixing is constructed in eight 
dimensions. The renormalization group functions are computed at one loop with
the special cases of $\Nc$~$=$~$2$ and $3$ treated separately. By computing the
critical exponents derived from these in the large $\Nf$ expansion at the 
Wilson-Fisher fixed point it is shown that the Lagrangian is in the same 
universality class as the two dimensional non-abelian Thirring model and 
Quantum Chromodynamics (QCD). As the eight dimensional Lagrangian contains new
quartic gluon operators  not present in four dimensional QCD, we compute in 
parallel the mixing matrix of {\em four} dimensional dimension $8$ operators in
pure Yang-Mills theory.  

\vspace{-17cm}
\hspace{13cm}
{\bf LTH 1146}

\newpage

\sect{Introduction.}

Non-abelian gauge theories are established as the core quantum field theories
which govern the particles of nature through the Standard Model. One sector,
which is known as Quantum Chromodynamics (QCD), describes the strong force 
between fundamental quarks and gluons which leads to the binding of these
quanta into the mesons and hadrons seen in nature. QCD has rather distinct
properties in comparison with the electroweak sector. For instance, at high
energy quarks and gluons become effectively free particles due to the property
of asymptotic freedom, \cite{1,2}. While this attribute is essential to 
developing a field theoretic formalism which allows us to extract meaningful
information from experimental data it has an implicit sense that at lower
energies quarks and gluons can never be treated as distinct particles in the 
same spirit as a free electron in Quantum Electrodynamics (QED) which is an
abelian gauge theory. The concept of a lack of low energy freedom is known as
colour confinement or infrared slavery in contradistinction to the virtual
freedom at ultraviolet scales. As it stands QCD has been studied at depth over
many years. One area where there has been significant progress recently is in
the evaluation of the fundamental renormalization group functions at very high
loop order. For instance, following the one loop discovery of asymptotic
freedom, \cite{1,2}, the two and three loop corrections to the $\beta$-function
appeared within a decade, \cite{3,4,5}. Progress to the four loop term followed
in the 1990's, \cite{6,7}, before a lull to the recent five loop explosion of 
all the renormalization group functions, \cite{8,9,10,11,12,13,14,15}. By this 
we mean the $\beta$-function was determined for the $SU(3)$ colour group in 
\cite{9} before this was extended to a general Lie group in \cite{10}. The 
supporting five loop renormalization group functions were determined in 
\cite{8,11,12,13,14,15}. While such multiloop QCD results are impressive in the 
extreme in the overall scheme of things having independent checks on such 
calculations is useful. The recent five loop QCD $\beta$-function of \cite{9} 
is relatively unique in this respect in that the independent computation of 
\cite{10} followed quickly. Ordinarily such a task requires as much human and 
computer resources as the initial breakthrough which are not always 
immediately available.

For QCD there is a parallel method of verifying part of the perturbative
series which is via the large $\Nf$ expansion where $\Nf$ is the number of
massless quarks. For instance, the QCD $\beta$-function was determined at
$O(1/\Nf)$ in \cite{16} which extended the QED result of \cite{17}. 
Subsequently the quark mass anomalous dimension was found at $O(1/\Nf^2)$ in 
\cite{18}. The $1/\Nf$ or large $\Nf$ expansion provides an alternative way of 
deducing certain coefficients in the perturbative series and the work of
\cite{16,18} extended the original method for spin-$0$ fields of \cite{19,20} 
to the spin-$1$ case. However, the formalism for the gauge theory context 
derives from a novel and elegant observation made in \cite{21}. In \cite{21} it
was shown that the non-abelian Thirring model (NATM) in the large $\Nf$ 
expansion is in the same universality class as QCD at the Wilson-Fisher fixed 
point in $d$-dimensions. While the non-abelian Thirring model is a 
non-renormalizable quantum field theory above two dimensions, within the large 
$\Nf$ expansion at its $d$-dimensional fixed point the $d$-dimensional critical
exponents contain information on the perturbative renormalization group 
functions of QCD. This has been verified by agreement with the latest set of 
five loop renormalization group functions, \cite{8,9,10,11,12,13,14,15}. The 
novel feature is the fact that in the non-abelian Thirring model there are 
{\em no} triple and quartic gluon self-interactions as is well known in QCD. 
These vertices effectively emerge at criticality within large $\Nf$ 
computations via $3$- and $4$-point quark loops, \cite{21}. More recently this 
property of critical equivalence has been studied in the simpler $O(N)$ scalar 
field theories where a similar phenomenon of higher dimensional theory vertices
are generated at criticality by triangle and box graphs. In more modern 
parlance this is known as ultraviolet completion. Indeed in the $O(N)$ 
nonlinear $\sigma$ model and $O(N)$ $\phi^4$ theory, the Wilson-Fisher fixed 
point equivalence in $2$~$<$~$d$~$<$~$4$ was extended to six dimensional $O(N)$ 
$\phi^3$ theory in \cite{22,23} and then beyond in \cite{24,25}. 

In light of this the six dimensional extension of the non-abelian Thirring
model and QCD equivalence was provided in \cite{26}. This involved a more 
intricate Lagrangian but the connection of the two loop renormalization group 
functions with the universal $d$-dimensional large $\Nf$ critical exponents was
verified. Again this reinforced the remarkable connection with the non-abelian 
Thirring model in that the six dimensional theory has quintic and sextic gluon 
self-interactions in addition to cubic and quartic structures which are the 
only ones present in four dimensions. While formally there are cubic and 
quartic interactions in both these dimensions the Feynman rules of the vertices
are different in each dimension. So the fact that the large $\Nf$ non-abelian 
Thirring model exponents encode information on the respective renormalization 
group functions is remarkable since it is not a gauge theory as such. Given 
this background it is therefore the purpose of this article to continue the 
tower of theories to the next link in the chain and construct the eight
dimensional non-abelian theory in what we will now term the non-abelian 
Thirring model universality class. This runs parallel to the six and eight 
dimensional extensions of QED, \cite{27,26}. The eight dimensional non-abelian 
theory has significantly more structure in its Lagrangian. For instance, there 
are seven independent quartic field strength operators in general as opposed to
two in the QED case, \cite{26}. Equally one has a higher power propagator for 
the gluon and Faddeev-Popov ghost fields which means evaluating Feynman 
integrals even at one loop becomes a significant task. Therefore in this 
article we concentrate on a full one loop renormalization of the field 
anomalous dimensions and all the $\beta$-functions. As such one can regard this
as proof of concept to launch a two loop computation from. The eight 
dimensional QED evaluation of \cite{26} was able to probe to two loops partly 
because of fewer interactions but also as a consequence of the Ward-Takahashi 
identity. 

A parallel reason for examining six and eight dimensional gauge theories rests 
in the connection to operators in lower dimensions. If one has the viewpoint of
an underlying universal theory residing at a fixed point in $d$-dimensions then
the gauge independent operators corresponding to the interactions of the higher
dimensional theory have dimensionless coupling constants in their respective 
critical dimensions. Below this dimension the coupling constant would become 
massive. Therefore they would equate to operators in the effective field theory
of the lower dimensional gauge theory. In \cite{26} it was noted that in the 
six dimensional extension of QCD the fully massive gluon propagator in the 
Landau gauge bore a remarkable qualitative similarity to the infrared behaviour
of the propagator as computed in the same gauge on the lattice but in 
{\em four} dimensions. While there was an observation in \cite{28,29} that the 
ultraviolet behaviour of a higher dimensional theory informs or models the 
infrared structure of a lower dimensional one, it would seem that an eight 
dimensional one could only relate to infrared fixed points in its six 
dimensional partner. However, given that dimension $8$ operators are of 
interest in four dimensional effective field theories of QCD having 
renormalization group function data in the eight dimensional non-abelian gauge 
theory for $SU(\Nc)$, where $\Nc$ is the number of colours, is an additional 
motivation for future studies. In four dimensions such dimension $8$ operators 
were studied in \cite{29} for Yang-Mills theories for the $SU(2)$ and $SU(3)$ 
colour groups. Here we extend the set and provide the one loop mixing matrix of
dimension $8$ operators in four dimensional $SU(\Nc)$ Yang-Mills theory. It 
will turn out that there are qualitative structural similarities between the 
matrix and the $\beta$-functions of the eight dimensional theory. 

The article is organized as follows. We discuss the construction of the eight
dimensional Lagrangian which will be in the same universality class as the
non-abelian Thirring model and QCD in the next section. The technology used to
renormalize the various $n$-point functions in this Lagrangian is discussed in
section $3$ before presenting the main results in section $4$. The connection
with the large $\Nf$ expansion of the critical exponents of the universality
class is checked in section $5$. In section $6$ we change tack and determine 
the mixing matrix of anomalous dimensions of dimension $8$ operators in four
dimensional Yang-Mills theory. Finally, concluding remarks are given in section
$7$.  

\sect{Background.}

As the first stage to constructing the eight dimensional version of QCD we
recall the corresponding Lagrangians of the lower dimensional cases. The four
dimensional Lagrangian is
\begin{equation}
L^{(4)} ~=~ -~ \frac{1}{4} G_{\mu\nu}^a G^{a \, \mu\nu} ~+~
i \bar{\psi}^{iI} \Dslash \psi^{iI} ~-~ 
\frac{1}{2\alpha} \left( \partial^\mu A^a_\mu \right)^2 ~-~
\bar{c}^a \left( \partial^\mu D_\mu c \right)^a
\label{lagqcd4}
\end{equation}
where we have included the canonical linear covariant gauge fixing term with
the associated Faddeev-Popov ghost. In (\ref{lagqcd4}) and throughout the gluon
field will be denoted by $A^a_\mu$, the quark field will be $\psi^{iI}$ and 
$c^a$ are the Faddeev-Popov ghost fields where $1$~$\leq$~$i$~$\leq$~$\Nf$,
$1$~$\leq$~$I$~$\leq$~$\NF$ and $1$~$\leq$~$a$~$\leq$~$\NA$. The parameters
$\Nf$, $\NA$ and $\NF$ correspond respectively to the number of (massless) 
quark flavours and the dimensions of the adjoint and fundamental 
representations of a general colour group. We use $\alpha$ as the linear 
covariant gauge parameter where $\alpha$~$=$~$0$ will correspond to the Landau 
gauge. To assist with the process of writing down the Lagrangians which are 
equivalent to (\ref{lagqcd4}) in higher dimensions one can regard 
(\ref{lagqcd4}) as being comprised of two parts. The first is the set of 
independent gauge invariant operators of dimension four built from the gluon 
and quark fields which have canonical dimensions of $1$ and $\frac{3}{2}$ in 
four dimensions. Then in order to be able to carry out explicit computations in
perturbation theory, for instance, one has to add in the appropriate gauge 
fixing term to ensure that a non-singular propagator can be constructed for the
gluon. This is the gauge fixing part of (\ref{lagqcd4}). From an operator point
of view this involves the independent gauge variant dimension four operators. 
By independent we mean those operators which are not related by linear 
combinations of total derivative operators. Given this the six dimensional 
extension of (\ref{lagqcd4}) was provided in \cite{24} based on similar work 
given in \cite{31}. With the increase in dimension the canonical dimension of 
the quark field is now $\frac{5}{2}$ which means that there are no quartic 
quark interactions. However, there are two independent gauge invariant gluonic 
operators which are apparent in the Lagrangian, \cite{24}, 
\begin{eqnarray}
L^{(6)} &=& -~ \frac{1}{4} \left( D_\mu G_{\nu\sigma}^a \right)
\left( D^\mu G^{a \, \nu\sigma} \right) ~+~
\frac{g_2}{6} f^{abc} G_{\mu\nu}^a \, G^{b \, \mu\sigma} \,
G^{c \,\nu}_{~~\,\sigma} \nonumber \\
&& -~ \frac{1}{2\alpha} \left( \partial_\mu \partial^\nu A^a_\nu \right)
\left( \partial^\mu \partial^\sigma A^a_\sigma \right) ~-~
\bar{c}^a \Box \left( \partial^\mu D_\mu c \right)^a ~+~
i \bar{\psi}^{iI} \Dslash \psi^{iI} 
\label{lagqcd6}
\end{eqnarray}
and mean that there are two coupling constants. Demonstrating the independence 
of the gluonic operators lies in part with the use of the Bianchi identity
\begin{equation}
D_\mu G^a_{\nu\sigma} ~+~ D_\nu G^a_{\sigma\mu} ~+~ 
D_\sigma G^a_{\mu\nu} ~=~ 0 ~.
\end{equation}
The remaining gauge invariant operator is the quark kinetic term wherein lies
the quark-gluon interaction which is the core interaction in the tower of
theories at the Wilson-Fisher fixed point. Throughout we will always denote the
usual gauge coupling constant by $g_1$ when there are one or more interactions.
The remaining part of (\ref{lagqcd6}) is completed with the dimension six 
linear covariant gauge fixing term which is the obvious extension of the four 
dimensional one. 

Equipped with this brief review of the construction of the dimension four and
six non-abelian gauge theories, the algorithm is now in place to proceed to
eight dimensions. In \cite{31,32} the renormalization of dimension eight 
operators in four dimensional Yang-Mills theory was considered and those
articles serve as the basis for the eight dimensional Lagrangian. As was 
discussed in \cite{32} there is only one independent dimension eight $2$-point
gauge invariant operator which therefore serves as the gluon kinetic term. 
Equally \cite{30,32} there are two independent dimension eight $3$-point gluon
operators. The new feature in eight dimensions, which derives from the fact 
that the gluon canonical dimension is unity, is that there will be quartic 
gluon field strength gauge invariant operators. The same property is present in
eight dimensional QED which was introduced in \cite{26} where there were 
several quartic photon self interactions. For the non-abelian case there is the
added complication of having to incorporate the colour group indices. The 
upshot is that one has to specify a particular colour group as it is not 
possible to have a finite set of quartic gluon opertors for a general Lie 
group, \cite{32}. Therefore we restrict ourselves to the $SU(\Nc)$ Lie group 
and recall relevant basic properties of this group needed for the Lagrangian. 
If $T^a$ is the Lie group generator then in $SU(\Nc)$ the product of two 
generators can be written as the linear combination 
\begin{equation}
T^a T^b ~=~ \frac{1}{2\Nc} \delta^{ab} ~+~ \frac{1}{2} d^{abc} T^c ~+~
\frac{i}{2} f^{abc} T^c
\end{equation}
where $d^{abc}$ is totally symmetric and the structure constants, $f^{abc}$,
are totally antisymmetric. Equally when we have to treat Feynman graphs with 
quarks the $SU(\Nc)$ relation  
\begin{equation}
T^a_{IJ} T^a_{KL} ~=~ \frac{1}{2} \left[ \delta_{IL} \delta_{KJ} ~-~
\frac{1}{\Nc} \delta_{IJ} \delta_{KL} \right]
\end{equation}
will be useful. To define gauge independent quartic gluon operators we 
introduce the rank $4$ colour tensors 
\begin{equation}
f_4^{abcd} ~\equiv~ f^{abe} f^{cde} ~~~,~~~ 
d_4^{abcd} ~\equiv~ d^{abe} d^{cde} 
\end{equation}
and then use the $SU(\Nc)$ relation between them, \cite{33},
\begin{equation}
f_4^{abcd} ~=~ \frac{2}{\Nc} \left( \delta^{ac} \delta^{bd} 
- \delta^{ad} \delta^{bc} \right) ~+~ d_4^{acbd} ~-~ d_4^{adbc} ~.
\end{equation} 
This in effect, \cite{33}, is the generalization of the relation between the
product of Levi-Civita tensors in $SU(2)$ to the colour groups $SU(\Nc)$ for
$\Nc$~$\geq$~$3$. It means that we use the tensor $d_4^{abcd}$ as the preferred
tensor of the gauge invariant operators. One reason for this is that 
$d_4^{abcd}$ is separately symmetric in the first or last pair of indices from
the full symmetry property of $d^{abc}$. Consequently there are eight gauge
independent quartic gluon operators in the eight dimensional extension of the
QCD Lagrangian leading to eleven independent coupling constants overall. The
full Lagrangian is 
\begin{eqnarray}
L^{(8)} &=& -~ \frac{1}{4} \left( D_\mu D_\nu G_{\sigma\rho}^a \right)
\left( D^\mu D^\nu G^{a \, \sigma\rho} \right) ~+~
\frac{g_2}{4} f^{abc} G_{\mu\nu}^a \, D^\mu G^{b \, \sigma\rho} \,
D^\nu G^c_{\sigma\rho} ~+~ i \bar{\psi}^{iI} \Dslash \psi^{iI} \nonumber \\
&& +~ \frac{g_3}{2} f^{abc} G_{\mu\nu}^a \, D_\sigma G^{b \, \mu\rho} \,
D^\sigma G^{c \,\nu}_{~~\,\rho} ~+~ 
g_4^2 G_{\mu \sigma}^a G^{a \, \mu \rho} G^{b \, \sigma \nu} G_{\rho \nu}^b ~+~ 
g_5^2 G_{\mu \sigma}^a G^{b \, \mu \rho} G^{b \, \sigma \nu} G_{\rho \nu}^a
\nonumber \\
&& +~ g_6^2 G_{\mu \sigma}^a G_{\nu \rho}^a G^{b \, \sigma \mu} 
G^{b \, \rho \nu} ~+~
g_7^2 G_{\mu \sigma}^a G_{\nu \rho}^b G^{a \, \sigma \mu} G^{b \, \rho \nu} ~+~
g_8^2 d_4^{a b c d} G_{\mu \sigma}^a G^{b \, \mu \sigma} G_{\nu \rho}^c
G^{d \, \nu \rho} \nonumber \\
&& +~ g_9^2 d_4^{a b c d} G_{\mu \sigma}^a G^{c \, \mu \rho} 
G^{b \, \nu \sigma} G_{\nu \rho}^d ~+~ 
g_{10}^2 d_4^{a c b d} G_{\mu \sigma}^a G^{b \, \mu \sigma} G_{\nu \rho}^c
G^{d \, \nu \rho} \nonumber \\
&& +~ g_{11}^2 d_4^{a d b c} G_{\mu \sigma}^a G^{c \, \mu \rho} 
G^{b \, \nu \sigma} G_{\nu \rho}^d ~-~ 
\frac{1}{2\alpha} \left( \partial_\mu \partial_\nu \partial^\sigma 
A^a_\sigma \right)
\left( \partial^\mu \partial^\nu \partial^\rho A^a_\rho \right) 
\nonumber \\ 
&& -~ \left( \Box \bar{c}^a \right) \left( \Box \partial^\mu D_\mu c \right)^a 
\label{lagqcd8}
\end{eqnarray}
where like (\ref{lagqcd4}) and (\ref{lagqcd6}) the dimension eight linear
covariant gauge fixing term is included. In addition the quark kinetic term is 
present and is equivalent to those in the lower dimensional Lagrangians which
therefore preserves the connection with the Wilson-Fisher fixed point and the
underlying universal theory which is accessible from the large $\Nf$ expansion.
While (\ref{lagqcd8}) represents the full $SU(\Nc)$ Lagrangian those for 
$\Nc$~$=$~$2$ and $3$ are smaller due to properties of the colour tensors. For 
instance, for the $SU(2)$ group $d^{abc}$~$=$~$0$. So for that group one has 
$g_8$~$=$~$g_9$~$=$~$g_{10}$~$=$~$g_{11}$~$=$~$0$. For $SU(3)$ 
$d^{abc}$~$\neq$~$0$ but $d^{abcd}$ satisfies  
\begin{equation}
d_4^{adbc} ~=~ -~ d_4^{abcd} ~-~ d_4^{acbd} ~+~ 
\frac{1}{3} \left[ \delta^{ab} \delta^{cd} + \delta^{ac} \delta^{bd} 
+ \delta^{ad} \delta^{bc} \right] ~.
\end{equation} 
This means that two of the operators involving $d^{abcd}$ are absent and within
our computations we have set $g_{10}$~$=$~$g_{11}$~$=$~$0$ for $SU(3)$. Finally
we note several useful $SU(\Nc)$ group identities, which we used within our 
graph evaluations, are, \cite{33},
\begin{equation}
d_4^{abcc} ~=~ 0 ~~~,~~~ 
d_4^{acbc} ~=~ \frac{[\Nc^2-4]}{\Nc} \delta^{ab} ~~~,~~~
d_4^{apbq} d_4^{cdpq} ~=~ \frac{[\Nc^2-12]}{2\Nc} d_4^{abcd} ~. 
\end{equation}
From the quadratic part of (\ref{lagqcd8}) in momentum space we find that the
gluon and ghost propagators are
\begin{eqnarray}
\langle A^a_\mu(p) A^b_\nu(-p) \rangle &=& -~ 
\frac{\delta^{ab}}{(p^2)^3} \left[ \eta_{\mu\nu} ~-~
(1 - \alpha) \frac{p_\mu p_\nu}{p^2} \right] \nonumber \\
\langle c^a(p) \bar{c}^b(-p) \rangle &=& -~ \frac{\delta^{ab}}{(p^2)^3}
\end{eqnarray}
which are formally the same as those in lower dimensions aside from the cubic
power of the overall factor. This is a similar feature to other eight 
dimensional theories and means that the evaluation of the Feynman graphs we
have to compute becomes exceedingly tedious.  

While we have constructed the most general non-abelian gauge theory based on a
simple Lie group in (\ref{lagqcd8}) this is in the case where there are no
masses present. The latter would not contribute to the renormalization group
functions at the Wilson-Fisher fixed point which is the main reason for not
considering them initially. However, one could view the presence of masses as
touching the lower dimensional operators which are allowed by power counting
renormalizability and which would be a staging point for connecting with the
other equivalent Lagrangians for this universality class. Therefore, budgeting
for non-zero masses (\ref{lagqcd8}) generalizes to  
\begin{eqnarray}
L_m^{(8)} &=& L^{(8)} + m_1 \bar{\psi}^{iI} \psi^{iI} ~-~ 
\frac{1}{4} m_2^2 \left( D_\mu G_{\nu\sigma}^a \right)
\left( D^\mu G^{a \, \nu\sigma} \right) ~-~ 
\frac{1}{2\alpha} m_3^2 \left( \partial_\mu \partial^\nu A^a_\nu \right)
\left( \partial^\mu \partial^\sigma A^a_\sigma \right) \nonumber \\
&& -~ m_3^2 \bar{c}^a \Box \left( \partial^\mu D_\mu c \right)^a ~-~ 
\frac{1}{4} m_4^4 G_{\mu\nu}^a G^{a \, \mu\nu} ~-~ 
\frac{1}{2\alpha} m_5^4 (\partial^\mu A^a_\mu)^2 ~-~
m_5^4 \bar{c}^a \left( \partial^\mu D_\mu c \right)^a \nonumber \\
&& -~ \frac{1}{2} m_6^6 A^a_\mu A^{a \, \mu} ~+~ m_6^6 \alpha \bar{c}^a c^a ~+~
\frac{1}{6} m_7^2 f^{abc} G_{\mu\nu}^a \, G^{b \, \mu\sigma} \,
G^{c \,\nu}_{~~\,\sigma} ~.
\label{lagqcd8m}
\end{eqnarray}
The additional terms fall into two classes which are operators which are gauge
invariant or not. In the latter case those operators are 
Becchi-Rouet-Stora-Tyutin (BRST) invariant. In particular it is evident that
the lower dimensional operators are a reflection of the Lagrangians of the
lower dimensional massless Lagrangians in the same universality class. In other
words in the critical dimension of the lower dimensional Lagrangians the masses
would correspond to coupling constants and hence be dimensionless in that 
spacetime. Implicit in (\ref{lagqcd8m}) is the assumption of locality. If one
ignored this and allowed for non-local operators then it is possible to 
construct a completely gauge invariant massive Lagrangian as discussed in
\cite{24}. The gluon and ghost propagators of (\ref{lagqcd8m}) have Stingl
forms, \cite{34}, since  
\begin{eqnarray}
\langle A^a_\mu(p) A^b_\nu(-p) \rangle &=& -~ 
\frac{\delta^{ab} P_{\mu\nu}(p)}
{[(p^2)^3 + m_2^2 (p^2)^2 + m_4^4 p^2 + m_6^6 ]} ~-~
\frac{\alpha \delta^{ab} L_{\mu\nu}(p)}{[(p^2)^3 + m_3^2 (p^2)^2 + m_5^2 p^2 
+ \alpha m_6^6 ]} \nonumber \\
\langle c^a(p) \bar{c}^b(-p) \rangle &=& -~
\frac{\delta^{ab}}{[(p^2)^3 + m_3^2 (p^2)^2 + m_5^4 p^2 + \alpha m_6^6 ]}
\end{eqnarray}
where
\begin{equation}
P_{\mu\nu}(p) ~=~ \eta_{\mu\nu} ~-~ \frac{p_\mu p_\nu}{p^2} ~~~,~~~
L_{\mu\nu}(p) ~=~ \frac{p_\mu p_\nu}{p^2} 
\end{equation}
are the respective transverse and longitudinal projection tensors. In this
formulation it is apparent that the pole structure of the Faddeev-Popov ghost
propagator matches that of the longitudinal part of the gluon. This ensures the
cancellation of unphysical degrees of freedom within computations with the 
massive Lagrangian.

\sect{Technical details.}

The task of renormalizing (\ref{lagqcd8}) requires several technical tools some
of which were applied to the determination of the two loop renormalization
group functions of $L^{(6)}$. However, with the presence of gauge independent
$4$-point operators built from the field strength the extraction of the 
$\beta$-functions of the respective coupling constants required a technique not
employed in \cite{24}. First, we note that we have constructed an automatic 
programme to renormalize the various $2$-, $3$- and $4$-point functions. The 
graphs contributing to each Green's function are generated using the 
{\sc Fortran} based package {\sc Qgraf}, \cite{35}. With the spinor, Lorentz
and colour group indices added to the electronic representation of the diagrams
each diagram is then passed to the integration routine specific to that 
particular $n$-point function. Once the divergences with respect to the 
regularization are known for each graph the full set is summed and the 
renormalization constants determined automatically without the use of the 
subtraction method but instead using the algorithm provided in \cite{36}. 
Briefly this is achieved by computing each Green's function as a function of 
the bare coupling constants and gauge parameter with their respective 
renormalized versions introduced by multiplicatively rescaling with the 
constant of proportionality being the renormalization constant. Specifically at
each loop order the renormalization constant associated with the Green's 
function is fixed by ensuring it is finite which determines the unknown 
counterterm at that order. Throughout this article we will consider only the 
$\MSbar$ scheme and regularize the theory using dimensional regularization 
where the spacetime dimension $d$ is set to $d$~$=$~$8$~$-$~$2\epsilon$ and 
$\epsilon$ is small. It acts as the regularization parameter. To handle the 
significant amounts of internal algebra of this whole process, use is made of 
the symbolic manipulation language {\sc Form}, \cite{37,38}. It is worth noting
that the renormalization of (\ref{lagqcd8}) involves $12$ independent 
parameters as well as colour and flavour parameters together with gluon and 
ghost propagators each of which have an exponent of $3$. This means there is a 
significant amount of integration to be performed, compared to four dimensional
QCD, for which {\sc Form} is the most efficient and practical tool for the 
task.

In order to construct the integration routine for each type of $n$-point 
function we follow what is now a well-established procedure which is the 
application of the integration by parts algorithm devised by Laporta, 
\cite{39}. To evaluate a Feynman graph it is first written as a sum of scalar 
integrals where scalar products of internal and external momenta are rewritten 
as combinations of the inverse propagators. For cases where there is no such
propagator in an integral, which is termed an irreducible, the basis of
propagators is extended or completed. It transpires that for each $n$-point
function at a particular loop order there is a small set of such independent
completions which are called integral families. These may or may not correspond
to an actual Feynman diagram topology. Irrespective of this it is the
mathematical representation of the integral family which is at the centre of
the Laporta method. One can determine a set of general algebraic relations 
between integrals in each family by integration by parts and Lorentz 
identities. The power of the Laporta algorithm is in realising that these 
relations can be solved algebraically in terms of a small set of basic or 
master Feynman integrals, \cite{39}. Thus if the $\epsilon$ expansion of these 
master integrals is known then all the Feynman integrals at that loop order can
be determined. In particular this includes the specific ones which comprise 
each of the graphs in the $n$-point functions of interest. There are various 
encodings of the Laporta algorithm available but we chose to use both versions 
of {\sc Reduze}, \cite{40,41}. While this outlines the general approach we used 
there are specific points which required attention. As we are renormalizing an 
{\em eight} dimensional Lagrangian we therefore need to have the master 
integrals in that dimension. Ordinarily the main focus in renormalization 
computations is four dimensions. However, we have not had to perform the 
explicit evaluation of master integrals by direct methods which is the normal 
way to determine their values. Instead we can exploit an elegant technique 
developed by Tarasov in \cite{42,43}. By considering the graph polynomial 
representation of a Feynman graph it is possible to relate a Feynman integral 
in $d$-dimensions in terms of a linear combination of the same integrals in 
$(d+2)$-dimensions. The latter, however, have several propagators with 
increased powers which is clearly necessary on dimensional grounds. This higher
dimensional set of integrals can be reduced to a linear combination of masters 
in the higher dimension. One of these will be the equivalent topology as the 
$d$-dimensional master with the remainder of the combination being masters with 
a fewer number of propagators, \cite{42,43}. As is the case in the Laporta 
algorithm some of these lower masters are integrals, such as simple bubble 
integrals, which are trivial to evaluate without using the Tarasov techniques. 
Therefore one can connect the more difficult to compute masters in 
$d$-dimensions with the unknown ones in $(d+2)$-dimensions. If the lower 
dimensional ones are available then the higher dimensional ones follow 
immediately. For our purposes we need to apply this connection twice since the 
various masters required are known in four dimensions. For instance, the 
$2$-point masters to four loops have been listed in \cite{44} while the
$3$-point masters for completely off-shell external legs were calculated to two
loops in \cite{45,46}. Also the one loop $4$-point box integral is known,
\cite{47}. Although we will not require the higher loop masters here it is 
worth noting what has been achieved over several years. 

This leads naturally to a brief discussion of the treatment of each set of
$n$-point functions separately. For the $2$-point functions and hence wave 
function renormalization constants we carried out the renormalization to two
loops. The main reason for this is that the double pole in $\epsilon$ of the
two loop renormalization constant is already pre-determined by the one loop
computation. Therefore this provides a partial check on the leading order
renormalization. For the $2$-point function we used the massless Lagrangian
and constructed the one and two loop masters by direct evaluation as these are
straightforward bubble integrals. By contrast for the $3$-point functions,
since nullifying an external leg leads to infrared issues we had to extend the
four dimensional off-shell massless master $3$-point function of \cite{44,46}
to eight dimensions using the Tarasov method, \cite{42,43}. For instance, if we
define the one loop triangle integral at the completely symmetric point by
\begin{equation}
I(\alpha,\beta,\gamma) ~=~ \int_k
\frac{1}{(k^2)^\alpha((k-p)^2)^\beta((k+q)^2)^\gamma}
\end{equation}
where $p$ and $q$ are the external momenta satisfying 
\begin{equation}
p^2 ~=~ q^2 ~=~ -~ \mu^2
\end{equation}
and $\int_k$~$=$~$d^dk/(2\pi)^d$ then
\begin{eqnarray}
\left. I(1,1,1) \right|_{d=8-2\epsilon} &=& -~ \mu^2 \left[ 
-~ \frac{1}{8\epsilon} - \frac{61}{144} - \frac{2\pi^2}{81} 
+ \frac{1}{27} \psi^\prime \left( \frac{1}{3} \right) \right. \nonumber \\
&& \left. ~~~~~~~\,+ \left[ \frac{1}{18} \psi^\prime \left( \frac{1}{3} \right) 
- \frac{895}{864} - \frac{23\pi^2}{864} 
- \frac{2}{3} s_3 \left( \frac{\pi}{6} \right)
+ \frac{35}{5832} \pi^3 \sqrt{3} \right. \right. \nonumber \\
&& \left. \left. ~~~~~~~~~~~~ + \frac{\pi}{216} \ln^2(3) \sqrt{3} \right] 
\epsilon ~+~ O(\epsilon^2) \right]
\label{mast111}
\end{eqnarray}
where $\psi(z)$~$=$~$\frac{d~}{dz} \ln \Gamma(z)$ and
\begin{equation}
s_n(z) ~=~ \frac{1}{\sqrt{3}} \Im \left[ \mbox{Li}_n \left(
\frac{e^{iz}}{\sqrt{3}} \right) \right]
\end{equation}
in terms of the polylogarithm function $\mbox{Li}_n(z)$. While only the simple
pole in $\epsilon$ is relevant for the renormalization of (\ref{lagqcd8}) we
have included the subsequent terms in the $\epsilon$ expansion for comparison
with the analogous lower dimensional masters. The finite part for instance is
directly correlated with the finite four dimensional master. The simple pole in
(\ref{mast111}) by contrast derives from the one loop bubble integrals which 
emerge in the Laporta reduction after the construction of the 
$(d+2)$-dimensional integrals from the $d$-dimensional master across two 
iterations. Equipped with (\ref{mast111}) the three coupling constants 
associated with the three independent $3$-point gluonic operators as well as
those of the quark and ghost vertices of (\ref{lagqcd8}) were renormalized 
using this strategy. For the latter vertices the quark-gluon vertex 
renormalization, for instance, determines the renormalization constant for 
$g_1$ which can be checked in the ghost-gluon vertex computation. For the 
remaining two couplings in this set, $g_2$ and $g_3$, their renormalization can
be determined from the gluon $3$-point vertex which provides a third check on 
the $\beta$-function of $g_1$. From examining the Feynman rule for the 
$3$-gluon vertex it can be seen that there are three independent tensor 
channels to provide three independent linear relations between the 
renormalization constants for these couplings.  

For the final part of the renormalization we have to extract the 
renormalization constants for the couplings associated with the purely quartic
operators of each eight dimensional Lagrangian. For this we used the vacuum
bubble expansion of \cite{48,49} as it was more efficient than constructing a 
large integration by parts database using {\sc Reduze}. This would be time
consuming to construct due to the high pole propagators for the gluon and
ghost. By contrast in the vacuum bubble expansion massless propagators are
recursively replaced by massive ones in such a way that the new propagators
eventually produce Feynman integrals which are ultraviolet finite. Hence by
Weinberg's theorem, \cite{50}, these do not contribute to the overall 
renormalization of the Green's function and so such terms can be neglected.
Subsequently the expansion terminates after a finite number of iterations. The
expansion is based on the exact identity, \cite{48,49},
\begin{equation}
\frac{1}{(k-p)^2} ~=~ \frac{1}{[k^2+m^2]} ~+~ 
\frac{2kp - p^2 + m^2}{(k-p)^2 [k^2+m^2]} ~.
\label{vacbubexp}
\end{equation}
The contribution to the overall degree of divergence of each of the numerator
pieces in the second term is less than that of the original propagator. In
addition the first term does not depend on the external momentum. So when all
such terms are collected within a Feynman integral it becomes a massive 
vacuum integral. Of course to produce the contributions which are purely vacuum
bubbles and contain the ultraviolet divergence of the Feynman graph the
identity has to be repeated sufficient times. Once this has been achieved a
simple Laporta reduction of one loop vacuum bubbles is constructed to reduce
the only one loop master vacuum bubble which is a simple standard integral in
eight dimensions. Another advantage of this approach is that the tensor 
structure arising from the external momenta together with the scalar products
of external momenta derived from (\ref{vacbubexp}) emerge relatively quickly.
In the summation of all the contributions to the gluon $4$-point function such
terms are central to disentangling the coupling constant renormalization 
constants for each of the independent quartic operators. A useful check on the
procedure is the absence of the parameter of the linear covariant gauge fixing 
in each of the coupling constant renormalizations in the three separate colour 
group computations we have to perform. 

\sect{Results.}

We turn now to the task of recording the results of our renormalization. First,
we have followed the conventions of previous analyses, \cite{24}, and note that
the renormalization of the parameter of the linear covariant gauge fixing is 
not independent of the gluon wave function renormalization in that  
\begin{equation}
\gamma_A(g_i) ~+~ \gamma_\alpha(g_i) ~=~ 0 ~.
\end{equation}
We have checked that this is true for all the $SU(\Nc)$ colour groups. For
$SU(2)$ the anomalous dimensions of the fields are  
\begin{eqnarray}
\left. \gamma_A^{SU(2)}(g_i) \right|_{\alpha = 0} 
&=& \left[ 24 \Nf g_1^2 + 871 g_1^2 - 4158 g_1 g_2 
- 1386 g_1 g_3 + 567 g_2^2 + 378 g_2 g_3 + 63 g_3^2 \right] \frac{1}{1680} 
\nonumber \\
&&
+ \left[ -~ 57594816 \Nf g_1^4 
- 2754788105 g_1^4 
+ 37417536 \Nf g_1^3 g_2 
+ 406217016 g_1^3 g_2
\right. \nonumber \\
&& ~~~ \left.
+~ 18601152 \Nf g_1^3 g_3 
+ 191078016 g_1^3 g_3
- 4398624 \Nf g_1^2 g_2^2 
\right. \nonumber \\
&& ~~~ \left.
-~ 1747949454 g_1^2 g_2^2
- 3900096 \Nf g_1^2 g_2 g_3 
- 2040796188 g_1^2 g_2 g_3
\right. \nonumber \\
&& ~~~ \left.
-~ 1053216 \Nf g_1^2 g_3^2 
- 261984978 g_1^2 g_3^2
+ 137535552 g_1^2 g_4^2
- 275071104 g_1^2 g_5^2
\right. \nonumber \\
&& ~~~ \left.
-~ 1124500608 g_1^2 g_6^2
+ 2249001216 g_1^2 g_7^2
+ 425614392 g_1 g_2^3
\right. \nonumber \\
&& ~~~ \left.
+~ 881618976 g_1 g_2^2 g_3
+ 500362128 g_1 g_2 g_3^2
+ 155288448 g_1 g_2 g_4^2
\right. \nonumber \\
&& ~~~ \left.
+~ 425614392 g_1 g_2^3
+ 881618976 g_1 g_2^2 g_3
- 310576896 g_1 g_2 g_5^2
\right. \nonumber \\
&& ~~~ \left.
+~ 234033408 g_1 g_2 g_6^2
+ 425614392 g_1 g_2^3
+ 881618976 g_1 g_2^2 g_3
\right. \nonumber \\
&& ~~~ \left.
-~ 468066816 g_1 g_2 g_7^2
+ 84640248 g_1 g_3^3
+ 425614392 g_1 g_2^3
\right. \nonumber \\
&& ~~~ \left.
+~ 881618976 g_1 g_2^2 g_3
+ 200785536 g_1 g_3 g_4^2
- 401571072 g_1 g_3 g_5^2
\right. \nonumber \\
&& ~~~ \left.
+~ 425614392 g_1 g_2^3
+ 881618976 g_1 g_2^2 g_3
- 21337344 g_1 g_3 g_6^2
\right. \nonumber \\
&& ~~~ \left.
+~ 42674688 g_1 g_3 g_7^2 
+ 425614392 g_1 g_2^3
+ 881618976 g_1 g_2^2 g_3
\right. \nonumber \\
&& ~~~ \left.
-~ 26643897 g_2^4 
- 87736068 g_2^3 g_3 
+ 425614392 g_1 g_2^3
+ 881618976 g_1 g_2^2 g_3
\right. \nonumber \\
&& ~~~ \left.
-~ 89488602 g_2^2 g_3^2 
- 52581312 g_2^2 g_4^2 
+ 425614392 g_1 g_2^3
+ 881618976 g_1 g_2^2 g_3
\right. \nonumber \\
&& ~~~ \left.
+~ 105162624 g_2^2 g_5^2 
+ 19813248 g_2^2 g_6^2 
+ 425614392 g_1 g_2^3
+ 881618976 g_1 g_2^2 g_3
\right. \nonumber \\
&& ~~~ \left.
-~ 39626496 g_2^2 g_7^2 
- 35913276 g_2 g_3^3 
+ 425614392 g_1 g_2^3
\right. \nonumber \\
&& ~~~ \left.
+~ 881618976 g_1 g_2^2 g_3
- 75696768 g_2 g_3 g_4^2 
+ 151393536 g_2 g_3 g_5^2 
\right. \nonumber \\
&& ~~~ \left.
+~ 425614392 g_1 g_2^3
+ 881618976 g_1 g_2^2 g_3
+ 40303872 g_2 g_3 g_6^2 
\right. \nonumber \\
&& ~~~ \left.
-~ 80607744 g_2 g_3 g_7^2 
+ 425614392 g_1 g_2^3
+ 881618976 g_1 g_2^2 g_3
\right. \nonumber \\
&& ~~~ \left.
-~ 5230701 g_3^4 
- 19389888 g_3^2 g_4^2 
+ 425614392 g_1 g_2^3
+ 881618976 g_1 g_2^2 g_3
\right. \nonumber \\
&& ~~~ \left.
+~ 38779776 g_3^2 g_5^2 
+ 11233152 g_3^2 g_6^2 
+ 425614392 g_1 g_2^3
+ 881618976 g_1 g_2^2 g_3
\right. \nonumber \\
&& ~~~ \left.
-~ 22466304 g_3^2 g_7^2 \right] \frac{1}{338688000} ~+~ O(g_i^6) 
\nonumber \\
\left. \gamma_c^{SU(2)}(g_i) \right|_{\alpha = 0} 
&=& -~ \frac{7}{24} g_1^2 \nonumber \\
&& +~ \left[ 12312 \Nf g_1^2 
-~ 3321487 g_1^2 
- 628614 g_1 g_2 
- 241878 g_1 g_3 
+ 77301 g_2^2 
\right. \nonumber \\
&& ~~~~ \left.
+~ 192654 g_2 g_3 
+ 108549 g_3^2 \right] \frac{g_1^2}{2419200} ~+~ O(g_i^6) 
\nonumber \\
\left. \gamma_\psi^{SU(2)}(g_i) \right|_{\alpha = 0} 
&=& \frac{7}{16} g_1^2 \nonumber \\ 
&& +~ \left[  
-~ 17352 \Nf g_1^2 
+ 3509752 g_1^2 
+ 1722294 g_1 g_2 
+ 973938 g_1 g_3 
- 196371 g_2^2 
\right. \nonumber \\
&& ~~~~ \left.
-~ 272034 g_2 g_3 
- 121779 g_3^2 \right] \frac{g_1^2}{1612800} ~+~ O(g_i^6) ~.
\end{eqnarray}
in the Landau gauge which is chosen for presentational reasons. The full
$\alpha$ dependent results are contained in the attached data file. One of the 
reasons for proceeding to two loops for this is as a check on the computation. 
The double pole in $\epsilon$ at two loops of the respective renormalization 
constants is not independent as it depends on the simple pole at one loop. We 
have verified that this is indeed the case in the explicit renormalization
constants for arbitrary $\alpha$. This checks the one loop coupling constant 
renormalization as well as the application of the Tarasov method, \cite{42,43},
to raise the four and six dimension massless two loop $2$-point master 
integrals to eight dimensions. The one loop $\beta$-functions are  
\begin{eqnarray}
\beta_1^{SU(2)}(g_i) &=&
\left[ 24 \Nf g_1^2 
- 109 g_1^2 
- 4158 g_1 g_2 
- 1386 g_1 g_3 
+ 567 g_2^2 
+ 378 g_2 g_3 
+ 63 g_3^2 \right] \frac{g_1}{3360} \nonumber \\
&& +~ O(g_i^5)  
\nonumber \\
\beta_2^{SU(2)}(g_i) &=&
\left[ 
- 272 \Nf g_1^3 
+ 32152 g_1^3 
+ 216 \Nf g_1^2 g_2 
+ 17919 g_1^2 g_2 
- 19908 g_1^2 g_3 
- 32634 g_1 g_2^2 
\right. \nonumber \\
&& ~ \left.
-~ 2646 g_1 g_2 g_3 
+ 3528 g_1 g_3^2 
+ 5103 g_2^3 
+ 2898 g_2^2 g_3 
- 441 g_2 g_3^2 
- 168 g_3^3 \right] \frac{1}{10080} \nonumber \\
&& +~ O(g_i^5)  
\nonumber \\
\beta_3^{SU(2)}(g_i) &=&
\left[ 
- 128 \Nf g_1^3 
- 18573 g_1^3 
+ 14889 g_1^2 g_2 
+ 36 \Nf g_1^2 g_3 
+ 8163 g_1^2 g_3 
- 2520 g_1 g_2^2 
\right. \nonumber \\
&& ~ \left.
-~ 7539 g_1 g_2 g_3 
- 777 g_1 g_3^2 
+ 5544 g_1 g_4^2 
- 11088 g_1 g_5^2 
- 3696 g_1 g_6^2 
+ 7392 g_1 g_7^2 
\right. \nonumber \\
&& ~ \left.
+~ 819 g_2^2 g_3 
+ 378 g_2 g_3^2 
- 1512 g_2 g_4^2 
+ 3024 g_2 g_5^2 
+ 1008 g_2 g_6^2 
- 2016 g_2 g_7^2 
\right. \nonumber \\
&& ~ \left.
-~ 21 g_3^3 
- 504 g_3 g_4^2 
+ 1008 g_3 g_5^2 
+ 336 g_3 g_6^2 
- 672 g_3 g_7^2 \right] \frac{1}{1680} ~+~ O(g_i^5)  
\nonumber \\
\beta_4^{SU(2)}(g_i) &=&
\left[ 800 \Nf g_1^4 
+ 73999 g_1^4 
- 82068 g_1^3 g_2 
- 48426 g_1^3 g_3 
+ 13734 g_1^2 g_2^2 
+ 12852 g_1^2 g_2 g_3 
\right. \nonumber \\
&& \left.
+~ 3360 g_1^2 g_3^2 
+ 1152 \Nf g_1^2 g_4^2 
- 89904 g_1^2 g_4^2 
- 32592 g_1^2 g_5^2 
- 113568 g_1^2 g_6^2 
\right. \nonumber \\
&& \left.
-~ 193536 g_1^2 g_7^2 
- 42 g_1 g_2^2 g_3 
+ 1008 g_1 g_2 g_3^2 
- 179424 g_1 g_2 g_4^2 
- 15456 g_1 g_2 g_5^2 
\right. \nonumber \\
&& \left.
-~ 2688 g_1 g_2 g_6^2 
+ 8064 g_1 g_2 g_7^2 
+ 2058 g_1 g_3^3 
- 6720 g_1 g_3 g_4^2 
- 7392 g_1 g_3 g_5^2 
\right. \nonumber \\
&& \left.
-~ 43008 g_1 g_3 g_6^2 
- 45696 g_1 g_3 g_7^2 
+ 27216 g_2^2 g_4^2 
+ 18144 g_2 g_3 g_4^2 
- 903 g_3^4 
\right. \nonumber \\
&& \left.
-~ 23184 g_3^2 g_4^2 
- 5712 g_3^2 g_5^2 
- 12768 g_3^2 g_6^2 
- 13440 g_3^2 g_7^2 
- 169344 g_4^4 
\right. \nonumber \\
&& \left.
-~ 188160 g_4^2 g_5^2 
- 177408 g_4^2 g_6^2 
- 139776 g_4^2 g_7^2 
- 124992 g_5^4 
- 145152 g_5^2 g_6^2 
\right. \nonumber \\
&& \left.
-~ 21504 g_5^2 g_7^2 
- 37632 g_6^4 
- 43008 g_6^2 g_7^2 
- 43008 g_7^4 \right] \frac{1}{40320} ~+~ O(g_i^6)
\nonumber \\
\beta_5^{SU(2)}(g_i) &=&
\left[ 
- 1192 \Nf g_1^4 
- 101355 g_1^4 
+ 84756 g_1^3 g_2 
+ 19194 g_1^3 g_3 
- 14070 g_1^2 g_2^2 
\right. \nonumber \\
&& \left.
-~ 16884 g_1^2 g_2 g_3 
+ 1848 g_1^2 g_3^2 
+ 52416 g_1^2 g_4^2 
+ 1152 \Nf g_1^2 g_5^2 
+ 16608 g_1^2 g_5^2 
\right. \nonumber \\
&& \left.
+~ 92064 g_1^2 g_6^2 
+ 193536 g_1^2 g_7^2 
+ 42 g_1 g_2^2 g_3 
+ 336 g_1 g_2 g_3^2 
- 12096 g_1 g_2 g_4^2 
\right. \nonumber \\
&& \left.
-~ 181440 g_1 g_2 g_5^2 
+ 8064 g_1 g_2 g_6^2 
- 8064 g_1 g_2 g_7^2 
+ 966 g_1 g_3^3 
- 11424 g_1 g_3 g_4^2 
\right. \nonumber \\
&& \left.
-~ 43008 g_1 g_3 g_5^2 
+ 75264 g_1 g_3 g_6^2 
+ 45696 g_1 g_3 g_7^2 
+ 27216 g_2^2 g_5^2 
\right. \nonumber \\
&& \left.
+~ 18144 g_2 g_3 g_5^2 
- 21 g_3^4 
+ 10080 g_3^2 g_4^2 
+ 3360 g_3^2 g_5^2 
+ 2016 g_3^2 g_6^2 
+ 13440 g_3^2 g_7^2 
\right. \nonumber \\
&& \left.
-~ 10752 g_4^4 
- 107520 g_4^2 g_5^2 
- 37632 g_4^2 g_6^2 
- 10752 g_4^2 g_7^2 
- 12096 g_5^4 
\right. \nonumber \\
&& \left.
-~ 26880 g_5^2 g_6^2 
- 129024 g_5^2 g_7^2 
- 26880 g_6^4 
- 43008 g_6^2 g_7^2 \right] \frac{1}{40320} ~+~ O(g_i^6)
\nonumber \\
\beta_6^{SU(2)}(g_i) &=&
\left[ 272 \Nf g_1^4 
- 248207 g_1^4 
+ 14742 g_1^3 g_2 
+ 134925 g_1^3 g_3 
+ 231 g_1^2 g_2^2 
- 7728 g_1^2 g_2 g_3 
\right. \nonumber \\
&& \left.
-~ 1323 g_1^2 g_3^2 
- 222432 g_1^2 g_4^2 
- 343392 g_1^2 g_5^2 
+ 2304 \Nf g_1^2 g_6^2 
- 1440480 g_1^2 g_6^2 
\right. \nonumber \\
&& \left.
-~ 228480 g_1^2 g_7^2 
+ 147 g_1 g_2^2 g_3 
- 4326 g_1 g_2 g_3^2 
+ 26880 g_1 g_2 g_4^2 
+ 48384 g_1 g_2 g_5^2 
\right. \nonumber \\
&& \left.
-~ 204288 g_1 g_2 g_6^2 
+ 2688 g_1 g_2 g_7^2 
- 4557 g_1 g_3^3 
+ 52416 g_1 g_3 g_4^2 
+ 118272 g_1 g_3 g_5^2 
\right. \nonumber \\
&& \left.
+~ 247296 g_1 g_3 g_6^2 
+ 34944 g_1 g_3 g_7^2 
+ 54432 g_2^2 g_6^2 
+ 36288 g_2 g_3 g_6^2 
- 42 g_3^4 
\right. \nonumber \\
&& \left.
+~ 3360 g_3^2 g_4^2 
+ 7392 g_3^2 g_5^2 
+ 50400 g_3^2 g_6^2 
- 21504 g_4^4 
- 80640 g_4^2 g_5^2 
\right. \nonumber \\
&& \left.
-~ 451584 g_4^2 g_6^2 
- 21504 g_4^2 g_7^2 
- 77952 g_5^4 
- 806400 g_5^2 g_6^2 
- 43008 g_5^2 g_7^2 
\right. \nonumber \\
&& \left.
-~ 1666560 g_6^4 
- 301056 g_6^2 g_7^2 \right] \frac{1}{80640} ~+~ O(g_i^6)
\nonumber \\
\beta_7^{SU(2)}(g_i) &=&
\left[ 8 \Nf g_1^4 
- 472989 g_1^4 
+ 154266 g_1^3 g_2 
+ 155883 g_1^3 g_3 
- 10647 g_1^2 g_2^2 
\right. \nonumber \\
&& \left.
-~ 31584 g_1^2 g_2 g_3 
+ 651 g_1^2 g_3^2 
+ 637056 g_1^2 g_4^2 
+ 480480 g_1^2 g_5^2 
+ 1704192 g_1^2 g_6^2 
\right. \nonumber \\
&& \left.
+~ 2304 \Nf g_1^2 g_7^2 
+ 3470496 g_1^2 g_7^2 
- 147 g_1 g_2^2 g_3 
- 1050 g_1 g_2 g_3^2 
- 72576 g_1 g_2 g_4^2 
\right. \nonumber \\
&& \left.
-~ 52416 g_1 g_2 g_5^2 
- 202944 g_1 g_2 g_6^2 
- 751296 g_1 g_2 g_7^2 
- 3339 g_1 g_3^3 
\right. \nonumber \\
&& \left.
-~ 124992 g_1 g_3 g_4^2 
- 81984 g_1 g_3 g_5^2 
- 307776 g_1 g_3 g_6^2 
- 651840 g_1 g_3 g_7^2 
\right. \nonumber \\
&& \left.
+~ 54432 g_2^2 g_7^2 
+ 36288 g_2 g_3 g_7^2 
- 42 g_3^4 
- 9408 g_3^2 g_4^2 
- 6048 g_3^2 g_5^2 
- 41664 g_3^2 g_6^2 
\right. \nonumber \\
&& \left.
- 74592 g_3^2 g_7^2 
- 91392 g_4^4 
- 80640 g_4^2 g_5^2 
- 408576 g_4^2 g_6^2 
- 1580544 g_4^2 g_7^2 
\right. \nonumber \\
&& \left.
-~ 5376 g_5^4 
- 64512 g_5^2 g_6^2 
- 913920 g_5^2 g_7^2 
- 118272 g_6^4 
- 3440640 g_6^2 g_7^2 
\right. \nonumber \\
&& \left.
-~ 4773888 g_7^4 \right] \frac{1}{80640} ~+~ O(g_i^6) ~. 
\end{eqnarray}
The main perturbative check on these expressions is the absence of the gauge
parameter. We computed the various $4$-point functions with non-zero $\alpha$
and verified that it cancelled in the final Green's function as it ought since
we are using the $\MSbar$ scheme.

The results for the case of $SU(3)$ are somewhat similar aside from the
additional two couplings. We have 
\begin{eqnarray}
\left. \gamma_A^{SU(3)}(g_i) \right|_{\alpha = 0}
&=& 
\left[ 16 g_1^2 \Nf 
+ 871 g_1^2 
- 4158 g_1 g_2 
- 1386 g_1 g_3 
+ 567 g_2^2 
+ 378 g_2 g_3 
+ 63 g_3^2 \right] \frac{1}{1120}
\nonumber \\
&& +~ \left[ 
-~ 110877632 g_1^4 \Nf 
- 8264364315 g_1^4 
+ 74835072 g_1^3 g_2 \Nf 
\right. \nonumber \\
&& ~~~~ \left.
+~ 1218651048 g_1^3 g_2 
+ 37202304 g_1^3 g_3 \Nf 
+ 573234048 g_1^3 g_3 
\right. \nonumber \\
&& ~~~~ \left.
-~ 8797248 g_1^2 g_2^2 \Nf 
- 5243848362 g_1^2 g_2^2 
- 7800192 g_1^2 g_2 g_3 \Nf 
\right. \nonumber \\
&& ~~~~ \left.
-~ 6122388564 g_1^2 g_2 g_3 
- 2106432 g_1^2 g_3^2 \Nf 
- 785954934 g_1^2 g_3^2 
\right. \nonumber \\
&& ~~~~ \left.
+~ 275071104 g_1^2 g_4^2 
- 550142208 g_1^2 g_5^2 
- 2249001216 g_1^2 g_6^2 
\right. \nonumber \\
&& ~~~~ \left.
+~ 4498002432 g_1^2 g_7^2 
+ 3748335360 g_1^2 g_8^2 
+ 229225920 g_1^2 g_9^2 
\right. \nonumber \\
&& ~~~~ \left.
+~ 1276843176 g_1 g_2^3 
+ 2644856928 g_1 g_2^2 g_3 
+ 1501086384 g_1 g_2 g_3^2 
\right. \nonumber \\
&& ~~~~ \left.
+~ 310576896 g_1 g_2 g_4^2 
- 621153792 g_1 g_2 g_5^2 
+ 468066816 g_1 g_2 g_6^2 
\right. \nonumber \\
&& ~~~~ \left.
-~ 936133632 g_1 g_2 g_7^2 
- 780111360 g_1 g_2 g_8^2 
+ 258814080 g_1 g_2 g_9^2 
\right. \nonumber \\
&& ~~~~ \left.
+~ 253920744 g_1 g_3^3 
+ 401571072 g_1 g_3 g_4^2 
- 803142144 g_1 g_3 g_5^2 
\right. \nonumber \\
&& ~~~~ \left.
-~ 42674688 g_1 g_3 g_6^2 
+ 85349376 g_1 g_3 g_7^2 
+ 71124480 g_1 g_3 g_8^2 
\right. \nonumber \\
&& ~~~~ \left.
+~ 334642560 g_1 g_3 g_9^2 
- 79931691 g_2^4 
- 263208204 g_2^3 g_3 
\right. \nonumber \\
&& ~~~~ \left.
-~ 268465806 g_2^2 g_3^2 
- 105162624 g_2^2 g_4^2 
+ 210325248 g_2^2 g_5^2 
\right. \nonumber \\
&& ~~~~ \left.
+~ 39626496 g_2^2 g_6^2 
- 79252992 g_2^2 g_7^2 
- 66044160 g_2^2 g_8^2 
\right. \nonumber \\
&& ~~~~ \left.
-~ 87635520 g_2^2 g_9^2 
- 107739828 g_2 g_3^3 
- 151393536 g_2 g_3 g_4^2 
\right. \nonumber \\
&& ~~~~ \left.
+~ 302787072 g_2 g_3 g_5^2 
+ 80607744 g_2 g_3 g_6^2 
- 161215488 g_2 g_3 g_7^2 
\right. \nonumber \\
&& ~~~~ \left.
-~ 134346240 g_2 g_3 g_8^2 
- 126161280 g_2 g_3 g_9^2 
- 15692103 g_3^4 
\right. \nonumber \\
&& ~~~~ \left.
-~ 38779776 g_3^2 g_4^2 
+ 77559552 g_3^2 g_5^2 
+ 22466304 g_3^2 g_6^2 
\right. \nonumber \\
&& ~~~~ \left.
-~ 44932608 g_3^2 g_7^2 
- 37443840 g_3^2 g_8^2 
- 32316480 g_3^2 g_9^2 \right] \frac{1}{451584000} ~+~ O(g_i^6) \nonumber \\
\left. \gamma_c^{SU(3)}(g_i) \right|_{\alpha = 0}
&=& 
-~ \frac{7}{16} g_1^2 \nonumber \\
&& +~ \left[ 8208 g_1^2 \Nf 
- 3321487 g_1^2 
- 628614 g_1 g_2 
- 241878 g_1 g_3 
+ 77301 g_2^2 
\right. \nonumber \\
&& ~~~ \left.
+~ 192654 g_2 g_3 
+ 108549 g_3^2 \right] \frac{g_1^2}{1075200} ~+~ O(g_i^6) \nonumber \\
\left. \gamma_\psi^{SU(3)}(g_i) \right|_{\alpha = 0}
&=& \frac{7}{9} g_1^2 \nonumber \\
&& +~ \left[ 
-~ 3856 g_1^2 \Nf 
+ 1147459 g_1^2 
+ 574098 g_1 g_2 
+ 324646 g_1 g_3 
- 65457 g_2^2 
\right. \nonumber \\
&& ~~~ \left.
-~ 90678 g_2 g_3 
- 40593 g_3^2 \right] 
\frac{g_1^2}{201600} ~+~ O(g_i^6) \nonumber \\
\beta_1^{SU(3)}(g_i) &=& 
\left[ 16 g_1^2 \Nf 
- 109 g_1^2 
- 4158 g_1 g_2 
- 1386 g_1 g_3 
+ 567 g_2^2 
+ 378 g_2 g_3 
+ 63 g_3^2 \right] \frac{g_1}{2240} \nonumber \\
&& +~ O(g_i^5) \nonumber \\
\beta_2^{SU(3)}(g_i) &=& 
\left[ 
-~ 544 g_1^3 \Nf 
+ 96456 g_1^3 
+ 432 g_1^2 g_2 \Nf 
+ 53757 g_1^2 g_2 
- 59724 g_1^2 g_3 
\right. \nonumber \\
&& ~ \left.
-~ 97902 g_1 g_2^2 
- 7938 g_1 g_2 g_3 
+ 10584 g_1 g_3^2 
+ 15309 g_2^3 
+ 8694 g_2^2 g_3 
\right. \nonumber \\
&& ~ \left.
-~ 1323 g_2 g_3^2 
- 504 g_3^3 \right] \frac{1}{20160} ~+~ O(g_i^5) \nonumber \\
\beta_3^{SU(3)}(g_i) &=& 
\left[ 
-~ 256 g_1^3 \Nf 
- 55719 g_1^3 
+ 44667 g_1^2 g_2 
+ 72 g_1^2 g_3 \Nf 
+ 24489 g_1^2 g_3 
- 7560 g_1 g_2^2 
\right. \nonumber \\
&& ~ \left.
-~ 22617 g_1 g_2 g_3 
- 2331 g_1 g_3^2 
+ 11088 g_1 g_4^2 
- 22176 g_1 g_5^2 
- 7392 g_1 g_6^2 
\right. \nonumber \\
&& ~ \left.
+~ 14784 g_1 g_7^2 
+ 12320 g_1 g_8^2 
+ 9240 g_1 g_9^2 
+ 2457 g_2^2 g_3 
+ 1134 g_2 g_3^2 
\right. \nonumber \\
&& ~ \left.
-~ 3024 g_2 g_4^2 
+ 6048 g_2 g_5^2 
+ 2016 g_2 g_6^2 
- 4032 g_2 g_7^2 
- 3360 g_2 g_8^2 
\right. \nonumber \\
&& ~ \left.
-~ 2520 g_2 g_9^2 
- 63 g_3^3 
- 1008 g_3 g_4^2 
+ 2016 g_3 g_5^2 
+ 672 g_3 g_6^2 
\right. \nonumber \\
&& ~ \left.
-~ 1344 g_3 g_7^2 
- 1120 g_3 g_8^2 
- 840 g_3 g_9^2 \right] \frac{1}{3360} ~+~ O(g_i^5) \nonumber \\
\beta_4^{SU(3)}(g_i) &=& 
\left[ 
- 784 g_1^4 \Nf 
- 61551 g_1^4 
+ 6048 g_1^3 g_2 
- 65772 g_1^3 g_3 
- 756 g_1^2 g_2^2 
- 9072 g_1^2 g_2 g_3 
\right. \nonumber \\
&& ~ \left.
+~ 11718 g_1^2 g_3^2 
+ 3456 g_1^2 g_4^2 \Nf 
- 168696 g_1^2 g_4^2 
- 208656 g_1^2 g_5^2 
- 417312 g_1^2 g_6^2 
\right. \nonumber \\
&& ~ \left.
-~ 245952 g_1^2 g_8^2 
- 85680 g_1^2 g_9^2 
+ 3024 g_1 g_2 g_3^2 
- 861840 g_1 g_2 g_4^2 
+ 6048 g_1 g_2 g_5^2 
\right. \nonumber \\
&& ~ \left.
+~ 12096 g_1 g_2 g_6^2 
+ 6048 g_1 g_2 g_9^2 
+ 6804 g_1 g_3^3 
- 81648 g_1 g_3 g_4^2 
- 30240 g_1 g_3 g_5^2 
\right. \nonumber \\
&& ~ \left.
-~ 60480 g_1 g_3 g_6^2 
- 88704 g_1 g_3 g_8^2 
+ 14112 g_1 g_3 g_9^2 
+ 122472 g_2^2 g_4^2 
\right. \nonumber \\
&& ~ \left.
+~ 81648 g_2 g_3 g_4^2 
- 2079 g_3^4 
- 58968 g_3^2 g_4^2 
- 27216 g_3^2 g_5^2 
- 54432 g_3^2 g_6^2 
\right. \nonumber \\
&& ~ \left.
-~ 20160 g_3^2 g_8^2 
- 17136 g_3^2 g_9^2 
- 870912 g_4^4 
- 806400 g_4^2 g_5^2 
- 1016064 g_4^2 g_6^2 
\right. \nonumber \\
&& ~ \left.
-~ 419328 g_4^2 g_7^2 
- 344064 g_4^2 g_8^2 
- 365568 g_4^2 g_9^2 
- 395136 g_5^4 
- 516096 g_5^2 g_6^2 
\right. \nonumber \\
&& ~ \left.
-~ 64512 g_5^2 g_7^2 
- 290304 g_5^2 g_8^2 
- 303744 g_5^2 g_9^2 
- 193536 g_6^4 
- 129024 g_6^2 g_7^2 
\right. \nonumber \\
&& ~ \left.
-~ 150528 g_6^2 g_8^2 
- 252672 g_6^2 g_9^2 
- 129024 g_7^4 
- 21504 g_7^2 g_9^2 
- 50176 g_8^4 
\right. \nonumber \\
&& ~ \left.
-~ 129024 g_8^2 g_9^2 
- 62720 g_9^4
\right] \frac{1}{120960} ~+~ O(g_i^6) \nonumber \\
\beta_5^{SU(3)}(g_i) &=& 
\left[ 
-~ 3576 g_1^4 \Nf 
- 445839 g_1^4 
+ 380394 g_1^3 g_2 
+ 97335 g_1^3 g_3 
- 63189 g_1^2 g_2^2 
\right. \nonumber \\
&& ~ \left.
-~ 74466 g_1^2 g_2 g_3 
+ 6363 g_1^2 g_3^2 
+ 157248 g_1^2 g_4^2 
+ 3456 g_1^2 g_5^2 \Nf 
+ 95400 g_1^2 g_5^2 
\right. \nonumber \\
&& ~ \left.
+~ 383040 g_1^2 g_6^2 
+ 580608 g_1^2 g_7^2 
+ 360864 g_1^2 g_8^2 
+ 88200 g_1^2 g_9^2 
+ 189 g_1 g_2^2 g_3 
\right. \nonumber \\
&& ~ \left.
+~ 1008 g_1 g_2 g_3^2 
- 36288 g_1 g_2 g_4^2 
- 841680 g_1 g_2 g_5^2 
+ 28224 g_1 g_2 g_6^2 
\right. \nonumber \\
&& ~ \left.
-~ 24192 g_1 g_2 g_7^2 
- 20160 g_1 g_2 g_8^2 
- 27216 g_1 g_2 g_9^2 
+ 3213 g_1 g_3^3 
\right. \nonumber \\
&& ~ \left.
-~ 34272 g_1 g_3 g_4^2 
- 194544 g_1 g_3 g_5^2 
+ 294336 g_1 g_3 g_6^2 
+ 137088 g_1 g_3 g_7^2 
\right. \nonumber \\
&& ~ \left.
+~ 69888 g_1 g_3 g_8^2 
- 21504 g_1 g_3 g_9^2 
+ 122472 g_2^2 g_5^2 
+ 81648 g_2 g_3 g_5^2 
+ 252 g_3^4 
\right. \nonumber \\
&& ~ \left.
+~ 30240 g_3^2 g_4^2 
+ 15624 g_3^2 g_5^2 
+ 8064 g_3^2 g_6^2 
+ 40320 g_3^2 g_7^2 
+ 23520 g_3^2 g_8^2 
\right. \nonumber \\
&& ~ \left.
+~ 16632 g_3^2 g_9^2 
- 32256 g_4^4 
- 322560 g_4^2 g_5^2 
- 112896 g_4^2 g_6^2 
- 32256 g_4^2 g_7^2 
\right. \nonumber \\
&& ~ \left.
-~ 10752 g_4^2 g_8^2 
- 21504 g_4^2 g_9^2 
- 56448 g_5^4 
- 161280 g_5^2 g_6^2 
- 387072 g_5^2 g_7^2 
\right. \nonumber \\
&& ~ \left.
-~ 172032 g_5^2 g_8^2 
- 118272 g_5^2 g_9^2 
- 161280 g_6^4 
- 129024 g_6^2 g_7^2 
- 129024 g_6^2 g_8^2 
\right. \nonumber \\
&& ~ \left.
-~ 59136 g_6^2 g_9^2 
- 10752 g_7^2 g_9^2 
- 7168 g_8^4 
+ 25088 g_8^2 g_9^2 
+ 10080 g_9^4 \right] \frac{1}{120960} \nonumber \\
&& +~ O(g_i^6) \nonumber \\
\beta_6^{SU(3)}(g_i) &=& 
\left[ 1632 g_1^4 \Nf 
- 1152069 g_1^4 
- 120834 g_1^3 g_2 
+ 778113 g_1^3 g_3 
+ 17703 g_1^2 g_2^2 
\right. \nonumber \\
&& ~ \left.
-~ 10584 g_1^2 g_2 g_3 
- 10899 g_1^2 g_3^2 
- 1334592 g_1^2 g_4^2 
- 1846656 g_1^2 g_5^2 
\right. \nonumber \\
&& ~ \left.
+~ 13824 g_1^2 g_6^2 \Nf 
- 8246880 g_1^2 g_6^2 
- 1370880 g_1^2 g_7^2 
- 747264 g_1^2 g_8^2 
\right. \nonumber \\
&& ~ \left.
-~ 391440 g_1^2 g_9^2 
+ 1323 g_1 g_2^2 g_3 
- 30870 g_1 g_2 g_3^2 
+ 161280 g_1 g_2 g_4^2 
\right. \nonumber \\
&& ~ \left.
+~ 298368 g_1 g_2 g_5^2 
- 2407104 g_1 g_2 g_6^2 
+ 16128 g_1 g_2 g_7^2 
- 26880 g_1 g_2 g_8^2 
\right. \nonumber \\
&& ~ \left.
+~ 55776 g_1 g_2 g_9^2 
- 29169 g_1 g_3^3 
+ 314496 g_1 g_3 g_4^2 
+ 846720 g_1 g_3 g_5^2 
\right. \nonumber \\
&& ~ \left.
+~ 1358784 g_1 g_3 g_6^2 
+ 209664 g_1 g_3 g_7^2 
+ 118272 g_1 g_3 g_8^2 
+ 139104 g_1 g_3 g_9^2 
\right. \nonumber \\
&& ~ \left.
+~ 489888 g_2^2 g_6^2 
+ 326592 g_2 g_3 g_6^2 
- 252 g_3^4 
+ 20160 g_3^2 g_4^2 
+ 48384 g_3^2 g_5^2 
\right. \nonumber \\
&& ~ \left.
+~ 328608 g_3^2 g_6^2 
+ 7728 g_3^2 g_9^2 
- 129024 g_4^4 
- 483840 g_4^2 g_5^2 
- 2709504 g_4^2 g_6^2 
\right. \nonumber \\
&& ~ \left.
-~ 129024 g_4^2 g_7^2 
- 43008 g_4^2 g_8^2 
- 86016 g_4^2 g_9^2 
- 548352 g_5^4 
- 5160960 g_5^2 g_6^2 
\right. \nonumber \\
&& ~ \left.
-~ 258048 g_5^2 g_7^2 
- 258048 g_5^2 g_8^2 
- 204288 g_5^2 g_9^2 
- 10321920 g_6^4 
- 1806336 g_6^2 g_7^2 
\right. \nonumber \\
&& ~ \left.
-~ 946176 g_6^2 g_8^2 
- 989184 g_6^2 g_9^2 
- 43008 g_7^2 g_9^2 
- 28672 g_8^4 
- 60928 g_8^2 g_9^2 
\right. \nonumber \\
&& ~ \left.
-~ 20160 g_9^4 \right] \frac{1}{483840} ~+~ O(g_i^6) \nonumber \\
\beta_7^{SU(3)}(g_i) &=& 
\left[ 608 g_1^4 \Nf 
- 5338695 g_1^4 
+ 1641906 g_1^3 g_2 
+ 1839159 g_1^3 g_3 
- 111447 g_1^2 g_2^2 
\right. \nonumber \\
&& ~ \left.
-~ 343224 g_1^2 g_2 g_3 
+ 4851 g_1^2 g_3^2 
+ 8797824 g_1^2 g_4^2 
+ 3673152 g_1^2 g_5^2 
\right. \nonumber \\
&& ~ \left.
+~ 11805696 g_1^2 g_6^2 
+ 27648 g_1^2 g_7^2 \Nf 
+ 59727168 g_1^2 g_7^2 
+ 1537536 g_1^2 g_8^2 
\right. \nonumber \\
&& ~ \left.
+~ 1832880 g_1^2 g_9^2 
- 1323 g_1 g_2^2 g_3 
- 17514 g_1 g_2 g_3^2 
- 983808 g_1 g_2 g_4^2 
\right. \nonumber \\
&& ~ \left.
-~ 395136 g_1 g_2 g_5^2 
- 1378944 g_1 g_2 g_6^2 
- 13491072 g_1 g_2 g_7^2 
- 53760 g_1 g_2 g_8^2 
\right. \nonumber \\
&& ~ \left.
-~ 213024 g_1 g_2 g_9^2 
- 41895 g_1 g_3^3 
- 1620864 g_1 g_3 g_4^2 
- 604800 g_1 g_3 g_5^2 
\right. \nonumber \\
&& ~ \left.
-~ 2072448 g_1 g_3 g_6^2 
- 11313792 g_1 g_3 g_7^2 
- 231168 g_1 g_3 g_8^2 
- 385056 g_1 g_3 g_9^2 
\right. \nonumber \\
&& ~ \left.
+~ 979776 g_2^2 g_7^2 
+ 653184 g_2 g_3 g_7^2 
- 504 g_3^4 
- 129024 g_3^2 g_4^2 
- 36288 g_3^2 g_5^2 
\right. \nonumber \\
&& ~ \left.
-~ 249984 g_3^2 g_6^2 
- 1342656 g_3^2 g_7^2 
- 25872 g_3^2 g_9^2 
- 2145024 g_4^4 
- 1128960 g_4^2 g_5^2 
\right. \nonumber \\
&& ~ \left.
-~ 5225472 g_4^2 g_6^2 
- 41545728 g_4^2 g_7^2 
- 258048 g_4^2 g_8^2 
- 838656 g_4^2 g_9^2 
- 64512 g_5^4 
\right. \nonumber \\
&& ~ \left.
-~ 774144 g_5^2 g_6^2 
- 14192640 g_5^2 g_7^2 
- 43008 g_5^2 g_8^2 
- 204288 g_5^2 g_9^2 
- 1419264 g_6^4 
\right. \nonumber \\
&& ~ \left.
-~ 47738880 g_6^2 g_7^2 
- 516096 g_6^2 g_8^2 
- 989184 g_6^2 g_9^2 
- 147603456 g_7^4 
\right. \nonumber \\
&& ~ \left.
-~ 4300800 g_7^2 g_8^2 
- 8644608 g_7^2 g_9^2 
- 172032 g_8^4 
- 60928 g_8^2 g_9^2 
\right. \nonumber \\
&& ~ \left.
-~ 82880 g_9^4 \right] \frac{1}{967680} ~+~ O(g_i^6) \nonumber \\
\beta_8^{SU(3)}(g_i) &=& 
\left[ 
-~ 512 g_1^4 \Nf 
- 2093313 g_1^4 
+ 881370 g_1^3 g_2 
+ 530523 g_1^3 g_3 
- 64575 g_1^2 g_2^2 
\right. \nonumber \\
&& ~ \left.
-~ 166320 g_1^2 g_2 g_3 
+ 7875 g_1^2 g_3^2 
+ 1334592 g_1^2 g_4^2 
+ 2487744 g_1^2 g_5^2 
\right. \nonumber \\
&& ~ \left.
+~ 9434880 g_1^2 g_6^2 
+ 1370880 g_1^2 g_7^2 
+ 9216 g_1^2 g_8^2 \Nf 
+ 10474176 g_1^2 g_8^2 
\right. \nonumber \\
&& ~ \left.
+~ 1021440 g_1^2 g_9^2 
- 1323 g_1 g_2^2 g_3 
+ 6678 g_1 g_2 g_3^2 
- 161280 g_1 g_2 g_4^2 
\right. \nonumber \\
&& ~ \left.
-~ 274176 g_1 g_2 g_5^2 
- 1137024 g_1 g_2 g_6^2 
- 16128 g_1 g_2 g_7^2 
- 3360000 g_1 g_2 g_8^2 
\right. \nonumber \\
&& ~ \left.
-~ 110208 g_1 g_2 g_9^2 
- 6363 g_1 g_3^3 
- 314496 g_1 g_3 g_4^2 
- 435456 g_1 g_3 g_5^2 
\right. \nonumber \\
&& ~ \left.
-~ 1733760 g_1 g_3 g_6^2 
- 209664 g_1 g_3 g_7^2 
- 2037504 g_1 g_3 g_8^2 
- 165312 g_1 g_3 g_9^2 
\right. \nonumber \\
&& ~ \left.
+~ 326592 g_2^2 g_8^2 
+ 217728 g_2 g_3 g_8^2 
- 126 g_3^4 
- 20160 g_3^2 g_4^2 
- 36288 g_3^2 g_5^2 
\right. \nonumber \\
&& ~ \left.
-~ 249984 g_3^2 g_6^2 
- 197568 g_3^2 g_8^2 
- 14784 g_3^2 g_9^2 
- 1720320 g_4^2 g_8^2 
- 75264 g_4^2 g_9^2 
\right. \nonumber \\
&& ~ \left.
-~ 2967552 g_5^2 g_8^2 
- 129024 g_5^2 g_9^2 
- 11956224 g_6^2 g_8^2 
- 666624 g_6^2 g_9^2 
\right. \nonumber \\
&& ~ \left.
-~ 1204224 g_7^2 g_8^2 
- 43008 g_7^2 g_9^2 
- 9619456 g_8^4 
- 1627136 g_8^2 g_9^2 
\right. \nonumber \\
&& ~ \left.
-~ 47488 g_9^4 \right] \frac{1}{322560} ~+~ O(g_i^6) \nonumber \\
\beta_9^{SU(3)}(g_i) &=& 
\left[ 6368 g_1^4 \Nf 
+ 830127 g_1^4 
- 754740 g_1^3 g_2 
- 260442 g_1^3 g_3 
+ 125622 g_1^2 g_2^2 
\right. \nonumber \\
&& ~ \left.
+~ 139860 g_1^2 g_2 g_3 
- 1008 g_1^2 g_3^2 
- 314496 g_1^2 g_4^2 
+ 82656 g_1^2 g_5^2 
- 124992 g_1^2 g_6^2 
\right. \nonumber \\
&& ~ \left.
-~ 1161216 g_1^2 g_7^2 
- 262080 g_1^2 g_8^2 
+ 2304 g_1^2 g_9^2 \Nf 
- 101376 g_1^2 g_9^2 
- 378 g_1 g_2^2 g_3 
\right. \nonumber \\
&& ~ \left.
+~ 1008 g_1 g_2 g_3^2 
+ 72576 g_1 g_2 g_4^2 
- 100800 g_1 g_2 g_5^2 
- 32256 g_1 g_2 g_6^2 
\right. \nonumber \\
&& ~ \left.
+~ 48384 g_1 g_2 g_7^2 
+ 48384 g_1 g_2 g_8^2 
- 536256 g_1 g_2 g_9^2 
+ 378 g_1 g_3^3 
+ 68544 g_1 g_3 g_4^2 
\right. \nonumber \\
&& ~ \left.
-~ 4032 g_1 g_3 g_5^2 
- 177408 g_1 g_3 g_6^2 
- 274176 g_1 g_3 g_7^2 
+ 86016 g_1 g_3 g_8^2 
\right. \nonumber \\
&& ~ \left.
-~ 63840 g_1 g_3 g_9^2 
+ 81648 g_2^2 g_9^2 
+ 54432 g_2 g_3 g_9^2 
- 2583 g_3^4 
- 60480 g_3^2 g_4^2 
\right. \nonumber \\
&& ~ \left.
+~ 2016 g_3^2 g_5^2 
- 4032 g_3^2 g_6^2 
- 80640 g_3^2 g_7^2 
- 22848 g_3^2 g_8^2 
- 30240 g_3^2 g_9^2 
\right. \nonumber \\
&& ~ \left.
-~ 107520 g_4^2 g_8^2 
- 172032 g_4^2 g_9^2 
+ 451584 g_5^2 g_8^2 
+ 177408 g_5^2 g_9^2 
+ 43008 g_6^2 g_8^2 
\right. \nonumber \\
&& ~ \left.
+~ 43008 g_6^2 g_9^2 
- 172032 g_7^2 g_8^2 
- 236544 g_7^2 g_9^2 
- 7168 g_8^4 
+ 28672 g_8^2 g_9^2 
\right. \nonumber \\
&& ~ \left.
-~ 32704 g_9^4 \right] \frac{1}{80640} ~+~ O(g_i^6) 
\end{eqnarray}
for the full set or renormalization group functions. 

The results for $SU(\Nc)$ are more involved partly because of the increase in 
the number of independent couplings but also because of the explicit $\Nc$ 
dependence. First, the Landau gauge field dimensions for $SU(\Nc)$ are
\begin{eqnarray}
\left. \gamma_A(g_i) \right|_{\alpha=0} &=& 
\left[ 871 \Nc g_1^2 + 48 \Nf g_1^2 - 4158 \Nc g_1 g_2 - 1386 \Nc g_1 g_3 
\right. \nonumber \\
&& \left. 
+~ 567 \Nc g_2^2 + 378 \Nc g_2 g_3 + 63 \Nc g_3^2 \right] 
\frac{1}{3360} ~+~ O(g_i^4)
\nonumber \\
\left. \gamma_c(g_i) \right|_{\alpha=0} &=&
- \frac{7}{48} g_1^2 \Nc \nonumber \\
&& + \left[ - 3321487 \Nc g_1^2 + 24624 \Nf g_1^2 - 628614 \Nc g_1 g_2
- 241878 \Nc g_1 g_3 + 77301 \Nc g_2^2 \right. \nonumber \\
&& \left. ~~~ + 192654 \Nc g_2 g_3 + 108549 \Nc g_3^2 \right] 
\frac{g_1^2 \Nc}{9676800} ~+~ O(g_i^6) \nonumber \\
\left. \gamma_\psi(g_i) \right|_{\alpha=0} &=&
\frac{7 [\Nc^2 - 1]}{24 \Nc} g_1^2 \nonumber \\
&& + \left[ 3388477 \Nc^4 g_1^2 - 34704 \Nc^3 \Nf g_1^2 - 2903377 \Nc^2 g_1^2 
+ 34704 \Nc \Nf g_1^2 - 485100 g_1^2 \right. \nonumber \\
&& ~~~~ \left.
+~ 1722294 \Nc^4 g_1 g_2 - 1722294 \Nc^2 g_1 g_2 + 973938 \Nc^4 g_1 g_3 
- 973938 \Nc^2 g_1 g_3
\right. \nonumber \\
&& ~~~~ \left.
-~ 196371 \Nc^4 g_2^2 + 196371 \Nc^2 g_2^2 - 272034 \Nc^4 g_2 g_3 
+ 272034 \Nc^2 g_2 g_3 
\right. \nonumber \\
&& ~~~~ \left.
-~ 121779 \Nc^4 g_3^2 + 121779 \Nc^2 g_3^2 \right] 
\frac{g_1^2}{4838400 \Nc^2} ~+~ O(g_i^6)
\label{rgesunc}
\end{eqnarray}
where we only present the two loop terms of the ghost and quark for
compactness. That for $\gamma_A(g_i)$ is given in the data file together with 
all the other renormalization group functions. For the $\beta$-functions we 
found  
\begin{eqnarray}
\beta_1(g_i) &=&
\frac{3}{320} \Nc g_1 g_3^2
+ \frac{9}{160} \Nc g_1 g_2 g_3
+ \frac{27}{320} \Nc g_1 g_2^2
- \frac{33}{160} \Nc g_1^2 g_3
- \frac{99}{160} \Nc g_1^2 g_2
- \frac{109}{6720} \Nc g_1^3
\nonumber \\
&&
+~ \frac{1}{140} \Nf g_1^3 
~+~ O(g_i^5) \nonumber \\
\beta_2(g_i) &=&
-~ \frac{1}{120} \Nc g_3^3
- \frac{7}{320} \Nc g_2 g_3^2
+ \frac{23}{160} \Nc g_2^2 g_3
+ \frac{81}{320} \Nc g_2^3
+ \frac{7}{40} \Nc g_1 g_3^2
- \frac{21}{160} \Nc g_1 g_2 g_3
\nonumber \\
&&
-~ \frac{259}{160} \Nc g_1 g_2^2
- \frac{79}{80} \Nc g_1^2 g_3
+ \frac{1991}{2240} \Nc g_1^2 g_2
+ \frac{4019}{2520} \Nc g_1^3
+ \frac{3}{140} \Nf g_1^2 g_2
\nonumber \\
&&
-~ \frac{17}{630} \Nf g_1^3 ~+~ O(g_i^5) \nonumber \\
\beta_3(g_i) &=&
- \frac{6}{5 \Nc} g_3 g_{11}^2
- \frac{2}{5 \Nc} g_3 g_{10}^2
+ \frac{3}{5 \Nc} g_3 g_9^2
+ \frac{4}{5 \Nc} g_3 g_8^2
- \frac{18}{5 \Nc} g_2 g_{11}^2
- \frac{6}{5 \Nc} g_2 g_{10}^2
+ \frac{9}{5 \Nc} g_2 g_9^2
\nonumber \\
&&
+~ \frac{12}{5 \Nc} g_2 g_8^2
+ \frac{66}{5 \Nc} g_1 g_{11}^2
+ \frac{22}{5 \Nc} g_1 g_{10}^2
- \frac{33}{5 \Nc} g_1 g_9^2
- \frac{44}{5 \Nc} g_1 g_8^2
- \frac{2}{5} g_3 g_7^2
+ \frac{1}{5} g_3 g_6^2
\nonumber \\
&&
+~ \frac{3}{5} g_3 g_5^2
- \frac{3}{10} g_3 g_4^2
- \frac{6}{5} g_2 g_7^2
+ \frac{3}{5} g_2 g_6^2
+ \frac{9}{5} g_2 g_5^2
- \frac{9}{10} g_2 g_4^2
+ \frac{22}{5} g_1 g_7^2
- \frac{11}{5} g_1 g_6^2
\nonumber \\
&&
-~ \frac{33}{5} g_1 g_5^2
+ \frac{33}{10} g_1 g_4^2
+ \frac{3}{10} \Nc g_3 g_{11}^2
+ \frac{1}{10} \Nc g_3 g_{10}^2
- \frac{3}{20} \Nc g_3 g_9^2
- \frac{1}{5} \Nc g_3 g_8^2
- \frac{1}{160} \Nc g_3^3
\nonumber \\
&&
+~ \frac{9}{10} \Nc g_2 g_{11}^2
+ \frac{3}{10} \Nc g_2 g_{10}^2
- \frac{9}{20} \Nc g_2 g_9^2
- \frac{3}{5} \Nc g_2 g_8^2
+ \frac{9}{80} \Nc g_2 g_3^2
+ \frac{39}{160} \Nc g_2^2 g_3
\nonumber \\
&&
-~ \frac{33}{10} \Nc g_1 g_{11}^2
- \frac{11}{10} \Nc g_1 g_{10}^2
+ \frac{33}{20} \Nc g_1 g_9^2
+ \frac{11}{5} \Nc g_1 g_8^2
- \frac{37}{160} \Nc g_1 g_3^2
- \frac{359}{160} \Nc g_1 g_2 g_3
\nonumber \\
&&
-~ \frac{3}{4} \Nc g_1 g_2^2
+ \frac{2721}{1120} \Nc g_1^2 g_3
+ \frac{709}{160} \Nc g_1^2 g_2
- \frac{6191}{1120} \Nc g_1^3
+ \frac{3}{140} \Nf g_1^2 g_3
- \frac{8}{105} \Nf g_1^3 
\nonumber \\
&&
+~ O(g_i^5) \nonumber \\
\beta_4(g_i) &=&
\frac{92}{5 \Nc^2} g_{11}^4
+ \frac{184}{15 \Nc^2} g_{10}^2 g_{11}^2
+ \frac{8}{5 \Nc^2} g_{10}^4
+ \frac{76}{5 \Nc^2} g_9^2 g_{11}^2
+ \frac{24}{5 \Nc^2} g_9^2 g_{10}^2
+ \frac{22}{3 \Nc^2} g_9^4
\nonumber \\
&&
+~ \frac{208}{15 \Nc^2} g_8^2 g_{11}^2
+ \frac{64}{15 \Nc^2} g_8^2 g_{10}^2
+ \frac{224}{15 \Nc^2} g_8^2 g_9^2
+ \frac{32}{5 \Nc^2} g_8^4
- \frac{1}{30 \Nc^2} g_3^2 g_{11}^2
+ \frac{1}{15 \Nc^2} g_3^2 g_{10}^2
\nonumber \\
&&
+~ \frac{1}{\Nc^2} g_3^2 g_9^2
+ \frac{4}{3 \Nc^2} g_3^2 g_8^2
+ \frac{1}{15 \Nc^2} g_1 g_3 g_{11}^2
+ \frac{44}{15 \Nc^2} g_1 g_3 g_{10}^2
- \frac{17}{15 \Nc^2} g_1 g_3 g_9^2
\nonumber \\
&&
+~ \frac{68}{15 \Nc^2} g_1 g_3 g_8^2
+ \frac{5}{3 \Nc^2} g_1 g_2 g_{11}^2
+ \frac{8}{15 \Nc^2} g_1 g_2 g_{10}^2
- \frac{6}{5 \Nc^2} g_1 g_2 g_9^2
- \frac{4}{5 \Nc^2} g_1 g_2 g_8^2
\nonumber \\
&&
-~ \frac{41}{30 \Nc^2} g_1^2 g_{11}^2
+ \frac{31}{15 \Nc^2} g_1^2 g_{10}^2
+ \frac{26}{5 \Nc^2} g_1^2 g_9^2
+ \frac{96}{5 \Nc^2} g_1^2 g_8^2
+ \frac{32}{5 \Nc} g_6^2 g_{11}^2
\nonumber \\
&&
+~ \frac{16}{15 \Nc} g_6^2 g_{10}^2
+ \frac{56}{15 \Nc} g_6^2 g_9^2
+ \frac{32}{15 \Nc} g_6^2 g_8^2
+ \frac{16}{5 \Nc} g_5^2 g_{11}^2
+ \frac{8}{15 \Nc} g_5^2 g_{10}^2
+ \frac{28}{15 \Nc} g_5^2 g_9^2
\nonumber \\
&&
+~ \frac{16}{15 \Nc} g_5^2 g_8^2
+ \frac{48}{5 \Nc} g_4^2 g_{11}^2
+ \frac{16}{5 \Nc} g_4^2 g_{10}^2
+ \frac{32}{5 \Nc} g_4^2 g_9^2
+ \frac{32}{5 \Nc} g_4^2 g_8^2
- \frac{2}{3 \Nc} g_3^2 g_7^2
\nonumber \\
&&
-~ \frac{1}{30 \Nc} g_3^2 g_6^2
+ \frac{1}{60 \Nc} g_3^2 g_5^2
- \frac{1}{2 \Nc} g_3^2 g_4^2
- \frac{34}{15 \Nc} g_1 g_3 g_7^2
- \frac{22}{15 \Nc} g_1 g_3 g_6^2
- \frac{1}{30 \Nc} g_1 g_3 g_5^2
\nonumber \\
&&
+~ \frac{17}{30 \Nc} g_1 g_3 g_4^2
+ \frac{2}{5 \Nc} g_1 g_2 g_7^2
- \frac{4}{15 \Nc} g_1 g_2 g_6^2
- \frac{5}{6 \Nc} g_1 g_2 g_5^2
+ \frac{3}{5 \Nc} g_1 g_2 g_4^2
\nonumber \\
&&
-~ \frac{48}{5 \Nc} g_1^2 g_7^2
- \frac{31}{30 \Nc} g_1^2 g_6^2
+ \frac{41}{60 \Nc} g_1^2 g_5^2
- \frac{13}{5 \Nc} g_1^2 g_4^2
- \frac{23}{5} g_{11}^4
- \frac{46}{15} g_{10}^2 g_{11}^2
- \frac{2}{5} g_{10}^4
\nonumber \\
&&
-~ \frac{19}{5} g_9^2 g_{11}^2
- \frac{6}{5} g_9^2 g_{10}^2
- \frac{11}{6} g_9^4
- \frac{52}{15} g_8^2 g_{11}^2
- \frac{16}{15} g_8^2 g_{10}^2
- \frac{56}{15} g_8^2 g_9^2
- \frac{8}{5} g_8^4
- \frac{16}{15} g_7^4
\nonumber \\
&&
-~ \frac{16}{15} g_6^2 g_7^2
- \frac{2}{5} g_6^4
- \frac{8}{15} g_5^2 g_7^2
- \frac{46}{15} g_5^2 g_6^2
- \frac{89}{30} g_5^4
- \frac{52}{15} g_4^2 g_7^2
- \frac{6}{5} g_4^2 g_6^2
- \frac{46}{15} g_4^2 g_5^2
\nonumber \\
&&
-~ \frac{9}{5} g_4^4
+ \frac{1}{120} g_3^2 g_{11}^2
- \frac{1}{60} g_3^2 g_{10}^2
- \frac{1}{4} g_3^2 g_9^2
- \frac{1}{3} g_3^2 g_8^2
- \frac{43}{1920} g_3^4
- \frac{1}{60} g_1 g_3 g_{11}^2
\nonumber \\
&&
-~ \frac{11}{15} g_1 g_3 g_{10}^2
+ \frac{17}{60} g_1 g_3 g_9^2
- \frac{17}{15} g_1 g_3 g_8^2
+ \frac{49}{960} g_1 g_3^3
- \frac{5}{12} g_1 g_2 g_{11}^2
- \frac{2}{15} g_1 g_2 g_{10}^2
\nonumber \\
&&
+~ \frac{3}{10} g_1 g_2 g_9^2
+ \frac{1}{5} g_1 g_2 g_8^2
+ \frac{1}{40} g_1 g_2 g_3^2
- \frac{1}{960} g_1 g_2^2 g_3
+ \frac{41}{120} g_1^2 g_{11}^2
- \frac{31}{60} g_1^2 g_{10}^2
\nonumber \\
&&
-~ \frac{13}{10} g_1^2 g_9^2
- \frac{24}{5} g_1^2 g_8^2
+ \frac{1}{12} g_1^2 g_3^2
+ \frac{51}{160} g_1^2 g_2 g_3
+ \frac{109}{320} g_1^2 g_2^2
- \frac{1153}{960} g_1^3 g_3
- \frac{977}{480} g_1^3 g_2
\nonumber \\
&&
+~ \frac{73999}{40320} g_1^4
- \frac{8}{5} \Nc g_6^2 g_{11}^2
- \frac{4}{15} \Nc g_6^2 g_{10}^2
- \frac{14}{15} \Nc g_6^2 g_9^2
- \frac{8}{15} \Nc g_6^2 g_8^2
- \frac{4}{5} \Nc g_5^2 g_{11}^2
\nonumber \\
&&
-~ \frac{2}{15} \Nc g_5^2 g_{10}^2
- \frac{7}{15} \Nc g_5^2 g_9^2
- \frac{4}{15} \Nc g_5^2 g_8^2
- \frac{12}{5} \Nc g_4^2 g_{11}^2
- \frac{4}{5} \Nc g_4^2 g_{10}^2
- \frac{8}{5} \Nc g_4^2 g_9^2
\nonumber \\
&&
-~ \frac{8}{5} \Nc g_4^2 g_8^2
- \frac{3}{20} \Nc g_3^2 g_6^2
- \frac{3}{40} \Nc g_3^2 g_5^2
- \frac{13}{80} \Nc g_3^2 g_4^2
+ \frac{9}{40} \Nc g_2 g_3 g_4^2
+ \frac{27}{80} \Nc g_2^2 g_4^2
\nonumber \\
&&
-~ \frac{1}{6} \Nc g_1 g_3 g_6^2
- \frac{1}{12} \Nc g_1 g_3 g_5^2
- \frac{9}{40} \Nc g_1 g_3 g_4^2
+ \frac{1}{30} \Nc g_1 g_2 g_6^2
+ \frac{1}{60} \Nc g_1 g_2 g_5^2
\nonumber \\
&&
-~ \frac{19}{8} \Nc g_1 g_2 g_4^2
- \frac{23}{20} \Nc g_1^2 g_6^2
- \frac{23}{40} \Nc g_1^2 g_5^2
- \frac{781}{1680} \Nc g_1^2 g_4^2
- \frac{2}{15} \Nc^2 g_6^4
- \frac{2}{15} \Nc^2 g_5^2 g_6^2
\nonumber \\
&&
-~ \frac{1}{30} \Nc^2 g_5^4
- \frac{4}{5} \Nc^2 g_4^2 g_6^2
- \frac{2}{5} \Nc^2 g_4^2 g_5^2
- \frac{3}{5} \Nc^2 g_4^4
+ \frac{5}{126 \Nc} \Nf g_1^4
+ \frac{1}{35} \Nf g_1^2 g_4^2 
\nonumber \\
&&
+~ O(g_i^6) \nonumber \\
\beta_5(g_i) &=&
\frac{4}{5 \Nc^2} g_{11}^4
+ \frac{8}{15 \Nc^2} g_{10}^2 g_{11}^2
+ \frac{8}{15 \Nc^2} g_{10}^4
+ \frac{52}{5 \Nc^2} g_9^2 g_{11}^2
+ \frac{56}{15 \Nc^2} g_9^2 g_{10}^2
+ \frac{6}{5 \Nc^2} g_9^4
\nonumber \\
&&
+~ \frac{176}{15 \Nc^2} g_8^2 g_{11}^2
+ \frac{64}{15 \Nc^2} g_8^2 g_{10}^2
+ \frac{32}{15 \Nc^2} g_8^2 g_9^2
+ \frac{32}{15 \Nc^2} g_8^4
+ \frac{1}{30 \Nc^2} g_3^2 g_{11}^2
- \frac{1}{15 \Nc^2} g_3^2 g_{10}^2
\nonumber \\
&&
-~ \frac{1}{\Nc^2} g_3^2 g_9^2
- \frac{4}{3 \Nc^2} g_3^2 g_8^2
- \frac{1}{15 \Nc^2} g_1 g_3 g_{11}^2
- \frac{44}{15 \Nc^2} g_1 g_3 g_{10}^2
+ \frac{17}{15 \Nc^2} g_1 g_3 g_9^2
\nonumber \\
&&
-~ \frac{68}{15 \Nc^2} g_1 g_3 g_8^2
- \frac{5}{3 \Nc^2} g_1 g_2 g_{11}^2
- \frac{8}{15 \Nc^2} g_1 g_2 g_{10}^2
+ \frac{6}{5 \Nc^2} g_1 g_2 g_9^2
+ \frac{4}{5 \Nc^2} g_1 g_2 g_8^2
\nonumber \\
&&
+~ \frac{41}{30 \Nc^2} g_1^2 g_{11}^2
- \frac{31}{15 \Nc^2} g_1^2 g_{10}^2
- \frac{26}{5 \Nc^2} g_1^2 g_9^2
- \frac{96}{5 \Nc^2} g_1^2 g_8^2
+ \frac{16}{15 \Nc} g_6^2 g_{10}^2
+ \frac{8}{15 \Nc} g_6^2 g_9^2
\nonumber \\
&&
+~ \frac{32}{15 \Nc} g_6^2 g_8^2
+ \frac{8}{15 \Nc} g_5^2 g_{10}^2
+ \frac{4}{15 \Nc} g_5^2 g_9^2
+ \frac{16}{15 \Nc} g_5^2 g_8^2
+ \frac{2}{3 \Nc} g_3^2 g_7^2
+ \frac{1}{30 \Nc} g_3^2 g_6^2
\nonumber \\
&&
-~ \frac{1}{60 \Nc} g_3^2 g_5^2
+ \frac{1}{2 \Nc} g_3^2 g_4^2
+ \frac{34}{15 \Nc} g_1 g_3 g_7^2
+ \frac{22}{15 \Nc} g_1 g_3 g_6^2
+ \frac{1}{30 \Nc} g_1 g_3 g_5^2
\nonumber \\
&&
-~ \frac{17}{30 \Nc} g_1 g_3 g_4^2
- \frac{2}{5 \Nc} g_1 g_2 g_7^2
+ \frac{4}{15 \Nc} g_1 g_2 g_6^2
+ \frac{5}{6 \Nc} g_1 g_2 g_5^2
- \frac{3}{5 \Nc} g_1 g_2 g_4^2
+ \frac{48}{5 \Nc} g_1^2 g_7^2
\nonumber \\
&&
+~ \frac{31}{30 \Nc} g_1^2 g_6^2
- \frac{41}{60 \Nc} g_1^2 g_5^2
+ \frac{13}{5 \Nc} g_1^2 g_4^2
- \frac{1}{5} g_{11}^4
- \frac{2}{15} g_{10}^2 g_{11}^2
- \frac{2}{15} g_{10}^4
- \frac{13}{5} g_9^2 g_{11}^2
\nonumber \\
&&
-~ \frac{14}{15} g_9^2 g_{10}^2
- \frac{3}{10} g_9^4
- \frac{44}{15} g_8^2 g_{11}^2
- \frac{16}{15} g_8^2 g_{10}^2
- \frac{8}{15} g_8^2 g_9^2
- \frac{8}{15} g_8^4
- \frac{16}{15} g_6^2 g_7^2
\nonumber \\
&&
-~ \frac{2}{15} g_6^4
- \frac{16}{5} g_5^2 g_7^2
- \frac{2}{15} g_5^2 g_6^2
- \frac{1}{6} g_5^4
- \frac{4}{15} g_4^2 g_7^2
- \frac{14}{15} g_4^2 g_6^2
- \frac{8}{3} g_4^2 g_5^2
- \frac{4}{15} g_4^4
\nonumber \\
&&
-~ \frac{1}{120} g_3^2 g_{11}^2
+ \frac{1}{60} g_3^2 g_{10}^2
+ \frac{1}{4} g_3^2 g_9^2
+ \frac{1}{3} g_3^2 g_8^2
- \frac{1}{1920} g_3^4
+ \frac{1}{60} g_1 g_3 g_{11}^2
+ \frac{11}{15} g_1 g_3 g_{10}^2
\nonumber \\
&&
-~ \frac{17}{60} g_1 g_3 g_9^2
+ \frac{17}{15} g_1 g_3 g_8^2
+ \frac{23}{960} g_1 g_3^3
+ \frac{5}{12} g_1 g_2 g_{11}^2
+ \frac{2}{15} g_1 g_2 g_{10}^2
- \frac{3}{10} g_1 g_2 g_9^2
\nonumber \\
&&
-~ \frac{1}{5} g_1 g_2 g_8^2
+ \frac{1}{120} g_1 g_2 g_3^2
+ \frac{1}{960} g_1 g_2^2 g_3
- \frac{41}{120} g_1^2 g_{11}^2
+ \frac{31}{60} g_1^2 g_{10}^2
+ \frac{13}{10} g_1^2 g_9^2
+ \frac{24}{5} g_1^2 g_8^2
\nonumber \\
&&
+~ \frac{11}{240} g_1^2 g_3^2
- \frac{67}{160} g_1^2 g_2 g_3
- \frac{67}{192} g_1^2 g_2^2
+ \frac{457}{960} g_1^3 g_3
+ \frac{1009}{480} g_1^3 g_2
- \frac{6757}{2688} g_1^4
- \frac{4}{15} \Nc g_6^2 g_{10}^2
\nonumber \\
&&
-~ \frac{2}{15} \Nc g_6^2 g_9^2
- \frac{8}{15} \Nc g_6^2 g_8^2
- \frac{2}{15} \Nc g_5^2 g_{10}^2
- \frac{1}{15} \Nc g_5^2 g_9^2
- \frac{4}{15} \Nc g_5^2 g_8^2
+ \frac{1}{60} \Nc g_3^2 g_6^2
\nonumber \\
&&
+~ \frac{11}{240} \Nc g_3^2 g_5^2
+ \frac{9}{40} \Nc g_2 g_3 g_5^2
+ \frac{27}{80} \Nc g_2^2 g_5^2
+ \frac{17}{30} \Nc g_1 g_3 g_6^2
- \frac{13}{24} \Nc g_1 g_3 g_5^2
\nonumber \\
&&
+ \frac{1}{30} \Nc g_1 g_2 g_6^2
- \frac{59}{24} \Nc g_1 g_2 g_5^2
+ \frac{53}{60} \Nc g_1^2 g_6^2
+ \frac{211}{560} \Nc g_1^2 g_5^2
- \frac{2}{15} \Nc^2 g_6^4
- \frac{2}{15} \Nc^2 g_5^2 g_6^2
\nonumber \\
&&
- \frac{1}{30} \Nc^2 g_5^4
- \frac{149}{2520 \Nc} \Nf g_1^4
+ \frac{1}{35} \Nf g_1^2 g_5^2 ~+~ O(g_i^6) \nonumber \\
\beta_6(g_i) &=&
\frac{52}{15 \Nc^2} g_{11}^4
+ \frac{568}{15 \Nc^2} g_{10}^2 g_{11}^2
+ \frac{1208}{15 \Nc^2} g_{10}^4
+ \frac{56}{15 \Nc^2} g_9^2 g_{11}^2
+ \frac{112}{5 \Nc^2} g_9^2 g_{10}^2
+ \frac{6}{5 \Nc^2} g_9^4
\nonumber \\
&&
+~ \frac{16}{15 \Nc^2} g_8^2 g_{11}^2
+ \frac{224}{15 \Nc^2} g_8^2 g_{10}^2
+ \frac{32}{15 \Nc^2} g_8^2 g_9^2
+ \frac{32}{15 \Nc^2} g_8^4
- \frac{3}{10 \Nc^2} g_3^2 g_{11}^2
- \frac{31}{15 \Nc^2} g_3^2 g_{10}^2
\nonumber \\
&&
-~ \frac{1}{6 \Nc^2} g_3^2 g_9^2
- \frac{18}{5 \Nc^2} g_1 g_3 g_{11}^2
- \frac{43}{3 \Nc^2} g_1 g_3 g_{10}^2
- \frac{13}{5 \Nc^2} g_1 g_3 g_9^2
- \frac{26}{15 \Nc^2} g_1 g_3 g_8^2
\nonumber \\
&&
-~ \frac{34}{15 \Nc^2} g_1 g_2 g_{11}^2
- \frac{47}{5 \Nc^2} g_1 g_2 g_{10}^2
- \frac{4}{3 \Nc^2} g_1 g_2 g_9^2
- \frac{2}{15 \Nc^2} g_1 g_2 g_8^2
+ \frac{617}{30 \Nc^2} g_1^2 g_{11}^2
\nonumber \\
&&
+~ \frac{78}{\Nc^2} g_1^2 g_{10}^2
+ \frac{331}{30 \Nc^2} g_1^2 g_9^2
+ \frac{34}{3 \Nc^2} g_1^2 g_8^2
+ \frac{16}{15 \Nc} g_6^2 g_{10}^2
+ \frac{8}{15 \Nc} g_6^2 g_9^2
+ \frac{32}{15 \Nc} g_6^2 g_8^2
\nonumber \\
&&
+~ \frac{8}{15 \Nc} g_5^2 g_{10}^2
+ \frac{4}{15 \Nc} g_5^2 g_9^2
+ \frac{16}{15 \Nc} g_5^2 g_8^2
+ \frac{31}{30 \Nc} g_3^2 g_6^2
+ \frac{3}{20 \Nc} g_3^2 g_5^2
+ \frac{1}{12 \Nc} g_3^2 g_4^2
\nonumber \\
&&
+~ \frac{13}{15 \Nc} g_1 g_3 g_7^2
+ \frac{43}{6 \Nc} g_1 g_3 g_6^2
+ \frac{9}{5 \Nc} g_1 g_3 g_5^2
+ \frac{13}{10 \Nc} g_1 g_3 g_4^2
+ \frac{1}{15 \Nc} g_1 g_2 g_7^2
\nonumber \\
&&
+~ \frac{47}{10 \Nc} g_1 g_2 g_6^2
+ \frac{17}{15 \Nc} g_1 g_2 g_5^2
+ \frac{2}{3 \Nc} g_1 g_2 g_4^2
- \frac{17}{3 \Nc} g_1^2 g_7^2
- \frac{39 \Nc} g_1^2 g_6^2
- \frac{617}{60 \Nc} g_1^2 g_5^2
\nonumber \\
&&
-~ \frac{331}{60 \Nc} g_1^2 g_4^2
- \frac{13}{15} g_{11}^4
- \frac{142}{15} g_{10}^2 g_{11}^2
- \frac{302}{15} g_{10}^4
- \frac{14}{15} g_9^2 g_{11}^2
- \frac{28}{5} g_9^2 g_{10}^2
- \frac{3}{10} g_9^4
\nonumber \\
&&
-~ \frac{4}{15} g_8^2 g_{11}^2
- \frac{56}{15} g_8^2 g_{10}^2
- \frac{8}{15} g_8^2 g_9^2
- \frac{8}{15} g_8^4
- \frac{56}{15} g_6^2 g_7^2
- \frac{302}{15} g_6^4
- \frac{8}{15} g_5^2 g_7^2
- \frac{142}{15} g_5^2 g_6^2
\nonumber \\
&&
-~ \frac{5}{6} g_5^4
- \frac{4}{15} g_4^2 g_7^2
- \frac{28}{5} g_4^2 g_6^2
- g_4^2 g_5^2
- \frac{4}{15} g_4^4
+ \frac{3}{40} g_3^2 g_{11}^2
+ \frac{31}{60} g_3^2 g_{10}^2
+ \frac{1}{24} g_3^2 g_9^2
\nonumber \\
&&
-~ \frac{1}{1920} g_3^4
+ \frac{9}{10} g_1 g_3 g_{11}^2
+ \frac{43}{12} g_1 g_3 g_{10}^2
+ \frac{13}{20} g_1 g_3 g_9^2
+ \frac{13}{30} g_1 g_3 g_8^2
- \frac{217}{3840} g_1 g_3^3
\nonumber \\
&&
+~ \frac{17}{30} g_1 g_2 g_{11}^2
+ \frac{47}{20} g_1 g_2 g_{10}^2
+ \frac{1}{3} g_1 g_2 g_9^2
+ \frac{1}{30} g_1 g_2 g_8^2
- \frac{103}{1920} g_1 g_2 g_3^2
+ \frac{7}{3840} g_1 g_2^2 g_3
\nonumber \\
&&
-~ \frac{617}{120} g_1^2 g_{11}^2
- \frac{39}{2} g_1^2 g_{10}^2
- \frac{331}{120} g_1^2 g_9^2
- \frac{17}{6} g_1^2 g_8^2
- \frac{21}{1280} g_1^2 g_3^2
- \frac{23}{240} g_1^2 g_2 g_3
\nonumber \\
&&
+~ \frac{11}{3840} g_1^2 g_2^2
+ \frac{1285}{768} g_1^3 g_3
+ \frac{117}{640} g_1^3 g_2
- \frac{248207}{80640} g_1^4
- \frac{4}{15} \Nc g_6^2 g_{10}^2
- \frac{2}{15} \Nc g_6^2 g_9^2
\nonumber \\
&&
-~ \frac{8}{15} \Nc g_6^2 g_8^2
- \frac{2}{15} \Nc g_5^2 g_{10}^2
- \frac{1}{15} \Nc g_5^2 g_9^2
- \frac{4}{15} \Nc g_5^2 g_8^2
+ \frac{13}{240} \Nc g_3^2 g_6^2
+ \frac{1}{120} \Nc g_3^2 g_5^2
\nonumber \\
&&
+~ \frac{9}{40} \Nc g_2 g_3 g_6^2
+ \frac{27}{80} \Nc g_2^2 g_6^2
- \frac{31}{120} \Nc g_1 g_3 g_6^2
+ \frac{17}{60} \Nc g_1 g_3 g_5^2
- \frac{293}{120} \Nc g_1 g_2 g_6^2
\nonumber \\
&&
+~ \frac{1}{60} \Nc g_1 g_2 g_5^2
+ \frac{275}{336} \Nc g_1^2 g_6^2
+ \frac{53}{120} \Nc g_1^2 g_5^2
- \frac{2}{15} \Nc^2 g_6^4
- \frac{2}{15} \Nc^2 g_5^2 g_6^2
- \frac{1}{30} \Nc^2 g_5^4
\nonumber \\
&&
+~ \frac{17}{2520 \Nc} \Nf g_1^4
+ \frac{1}{35} \Nf g_1^2 g_6^2 ~+~ O(g_i^6) \nonumber \\
\beta_7(g_i) &=&
\frac{52}{15 \Nc^2} g_{11}^4
+ \frac{568}{15 \Nc^2} g_{10}^2 g_{11}^2
+ \frac{1208}{15 \Nc^2} g_{10}^4
+ \frac{56}{15 \Nc^2} g_9^2 g_{11}^2
+ \frac{328}{15 \Nc^2} g_9^2 g_{10}^2
+ \frac{16}{15 \Nc^2} g_9^4
\nonumber \\
&&
+~ \frac{16}{15 \Nc^2} g_8^2 g_{11}^2
+ \frac{32}{3 \Nc^2} g_8^2 g_{10}^2
+ \frac{16}{15 \Nc^2} g_8^2 g_9^2
+ \frac{32}{15 \Nc^2} g_8^4
+ \frac{3}{10 \Nc^2} g_3^2 g_{11}^2
+ \frac{31}{15 \Nc^2} g_3^2 g_{10}^2
\nonumber \\
&&
+~ \frac{1}{6 \Nc^2} g_3^2 g_9^2
+ \frac{18}{5 \Nc^2} g_1 g_3 g_{11}^2
+ \frac{43}{3 \Nc^2} g_1 g_3 g_{10}^2
+ \frac{13}{5 \Nc^2} g_1 g_3 g_9^2
+ \frac{26}{15 \Nc^2} g_1 g_3 g_8^2
\nonumber \\
&&
+~ \frac{34}{15 \Nc^2} g_1 g_2 g_{11}^2
+ \frac{47}{5 \Nc^2} g_1 g_2 g_{10}^2
+ \frac{4}{3 \Nc^2} g_1 g_2 g_9^2
+ \frac{2}{15 \Nc^2} g_1 g_2 g_8^2
- \frac{617}{30 \Nc^2} g_1^2 g_{11}^2
\nonumber \\
&&
-~ \frac{78}{\Nc^2} g_1^2 g_{10}^2
- \frac{331}{30 \Nc^2} g_1^2 g_9^2
- \frac{34}{3 \Nc^2} g_1^2 g_8^2
+ \frac{112}{3 \Nc} g_7^2 g_{11}^2
+ \frac{464}{3 \Nc} g_7^2 g_{10}^2
+ \frac{64}{3 \Nc} g_7^2 g_9^2
\nonumber \\
&&
+~ \frac{32}{3 \Nc} g_7^2 g_8^2
+ \frac{8}{15 \Nc} g_6^2 g_{11}^2
+ \frac{16}{3 \Nc} g_6^2 g_{10}^2
+ \frac{4}{15 \Nc} g_6^2 g_9^2
+ \frac{4}{15 \Nc} g_5^2 g_{11}^2
+ \frac{8}{3 \Nc} g_5^2 g_{10}^2
\nonumber \\
&&
+~ \frac{2}{15 \Nc} g_5^2 g_9^2
+ \frac{52}{15 \Nc} g_4^2 g_{11}^2
+ \frac{284}{15 \Nc} g_4^2 g_{10}^2
+ \frac{28}{15 \Nc} g_4^2 g_9^2
+ \frac{8}{15 \Nc} g_4^2 g_8^2
- \frac{31}{30 \Nc} g_3^2 g_6^2
\nonumber \\
&&
-~ \frac{3}{20 \Nc} g_3^2 g_5^2
- \frac{1}{12 \Nc} g_3^2 g_4^2
- \frac{13}{15 \Nc} g_1 g_3 g_7^2
- \frac{43}{6 \Nc} g_1 g_3 g_6^2
- \frac{9}{5 \Nc} g_1 g_3 g_5^2
- \frac{13}{10 \Nc} g_1 g_3 g_4^2
\nonumber \\
&&
-~ \frac{1}{15 \Nc} g_1 g_2 g_7^2
- \frac{47}{10 \Nc} g_1 g_2 g_6^2
- \frac{17}{15 \Nc} g_1 g_2 g_5^2
- \frac{2}{3 \Nc} g_1 g_2 g_4^2
+ \frac{17}{3 \Nc} g_1^2 g_7^2
+ \frac{39}{\Nc} g_1^2 g_6^2
\nonumber \\
&&
+~ \frac{617}{60 \Nc} g_1^2 g_5^2
+ \frac{331}{60 \Nc} g_1^2 g_4^2
- \frac{13}{15} g_{11}^4
- \frac{142}{15} g_{10}^2 g_{11}^2
- \frac{302}{15} g_{10}^4
- \frac{14}{15} g_9^2 g_{11}^2
- \frac{82}{15} g_9^2 g_{10}^2
\nonumber \\
&&
-~ \frac{4}{15} g_9^4
- \frac{4}{15} g_8^2 g_{11}^2
- \frac{8}{3} g_8^2 g_{10}^2
- \frac{4}{15} g_8^2 g_9^2
- \frac{8}{15} g_8^4
+ \frac{232}{15} g_7^4
- \frac{112}{3} g_6^2 g_7^2
- \frac{22}{15} g_6^4
\nonumber \\
&&
-~ \frac{26}{3} g_5^2 g_7^2
- \frac{4}{5} g_5^2 g_6^2
- \frac{1}{15} g_5^4
- \frac{14}{15} g_4^2 g_7^2
- \frac{24}{5} g_4^2 g_6^2
- \frac{13}{15} g_4^2 g_5^2
- \frac{4}{15} g_4^4
- \frac{3}{40} g_3^2 g_{11}^2
\nonumber \\
&&
-~ \frac{31}{60} g_3^2 g_{10}^2
- \frac{1}{24} g_3^2 g_9^2
- \frac{1}{1920} g_3^4
- \frac{9}{10} g_1 g_3 g_{11}^2
- \frac{43}{12} g_1 g_3 g_{10}^2
- \frac{13}{20} g_1 g_3 g_9^2
- \frac{13}{30} g_1 g_3 g_8^2
\nonumber \\
&&
-~ \frac{53}{1280} g_1 g_3^3
- \frac{17}{30} g_1 g_2 g_{11}^2
- \frac{47}{20} g_1 g_2 g_{10}^2
- \frac{1}{3} g_1 g_2 g_9^2
- \frac{1}{30} g_1 g_2 g_8^2
- \frac{5}{384} g_1 g_2 g_3^2
\nonumber \\
&&
-~ \frac{7}{3840} g_1 g_2^2 g_3
+ \frac{617}{120} g_1^2 g_{11}^2
+ \frac{39}{2} g_1^2 g_{10}^2
+ \frac{331}{120} g_1^2 g_9^2
+ \frac{17}{6} g_1^2 g_8^2
+ \frac{31}{3840} g_1^2 g_3^2
\nonumber \\
&&
-~ \frac{47}{120} g_1^2 g_2 g_3
- \frac{169}{1280} g_1^2 g_2^2
+ \frac{7423}{3840} g_1^3 g_3
+ \frac{3673}{1920} g_1^3 g_2
- \frac{157663}{26880} g_1^4
- \frac{28}{3} \Nc g_7^2 g_{11}^2
\nonumber \\
&&
-~ \frac{116}{3} \Nc g_7^2 g_{10}^2
- \frac{16}{3} \Nc g_7^2 g_9^2
- \frac{8}{3} \Nc g_7^2 g_8^2
- \frac{2}{15} \Nc g_6^2 g_{11}^2
- \frac{4}{3} \Nc g_6^2 g_{10}^2
- \frac{1}{15} \Nc g_6^2 g_9^2
\nonumber \\
&&
-~ \frac{1}{15} \Nc g_5^2 g_{11}^2
- \frac{2}{3} \Nc g_5^2 g_{10}^2
- \frac{1}{30} \Nc g_5^2 g_9^2
- \frac{13}{15} \Nc g_4^2 g_{11}^2
- \frac{71}{15} \Nc g_4^2 g_{10}^2
- \frac{7}{15} \Nc g_4^2 g_9^2
\nonumber \\
&&
-~ \frac{2}{15} \Nc g_4^2 g_8^2
- \frac{37}{80} \Nc g_3^2 g_7^2
- \frac{3}{80} \Nc g_3^2 g_4^2
+ \frac{9}{40} \Nc g_2 g_3 g_7^2
+ \frac{27}{80} \Nc g_2^2 g_7^2
- \frac{153}{40} \Nc g_1 g_3 g_7^2
\nonumber \\
&&
-~ \frac{7}{60} \Nc g_1 g_3 g_6^2
- \frac{7}{120} \Nc g_1 g_3 g_5^2
- \frac{9}{20} \Nc g_1 g_3 g_4^2
- \frac{557}{120} \Nc g_1 g_2 g_7^2
- \frac{1}{12} \Nc g_1 g_2 g_6^2
\nonumber \\
&&
-~ \frac{1}{24} \Nc g_1 g_2 g_5^2
- \frac{17}{60} \Nc g_1 g_2 g_4^2
+ \frac{11257}{560} \Nc g_1^2 g_7^2
+ \frac{49}{60} \Nc g_1^2 g_6^2
+ \frac{49}{120} \Nc g_1^2 g_5^2
\nonumber \\
&&
+~ \frac{617}{240} \Nc g_1^2 g_4^2
- \frac{56}{3} \Nc^2 g_7^4
- \frac{4}{3} \Nc^2 g_6^2 g_7^2
- \frac{2}{3} \Nc^2 g_5^2 g_7^2
- \frac{14}{3} \Nc^2 g_4^2 g_7^2
- \frac{1}{15} \Nc^2 g_4^2 g_6^2
\nonumber \\
&&
-~ \frac{1}{30} \Nc^2 g_4^2 g_5^2
- \frac{13}{60} \Nc^2 g_4^4
+ \frac{1}{5040 \Nc} \Nf g_1^4
+ \frac{1}{35} \Nf g_1^2 g_7^2 ~+~ O(g_i^6) \nonumber \\
\beta_8(g_i) &=&
\frac{58}{15 \Nc} g_{11}^4
+ \frac{128}{3 \Nc} g_{10}^2 g_{11}^2
+ \frac{1328}{15 \Nc} g_{10}^4
+ \frac{139}{15 \Nc} g_9^2 g_{11}^2
+ \frac{764}{15 \Nc} g_9^2 g_{10}^2
+ \frac{71}{15 \Nc} g_9^4
\nonumber \\
&&
+~ \frac{896}{15 \Nc} g_8^2 g_{11}^2
+ \frac{248 \Nc} g_8^2 g_{10}^2
+ \frac{788}{15 \Nc} g_8^2 g_9^2
+ \frac{464}{5 \Nc} g_8^4
+ \frac{3}{20 \Nc} g_3^2 g_{11}^2
+ \frac{31}{30 \Nc} g_3^2 g_{10}^2
\nonumber \\
&&
+~ \frac{1}{12 \Nc} g_3^2 g_9^2
+ \frac{9}{5 \Nc} g_1 g_3 g_{11}^2
+ \frac{43}{6 \Nc} g_1 g_3 g_{10}^2
+ \frac{13}{10 \Nc} g_1 g_3 g_9^2
+ \frac{13}{15 \Nc} g_1 g_3 g_8^2
\nonumber \\
&&
+~ \frac{17}{15 \Nc} g_1 g_2 g_{11}^2
+ \frac{47}{10 \Nc} g_1 g_2 g_{10}^2
+ \frac{2}{3 \Nc} g_1 g_2 g_9^2
+ \frac{1}{15 \Nc} g_1 g_2 g_8^2
- \frac{617}{60 \Nc} g_1^2 g_{11}^2
- \frac{39}{\Nc} g_1^2 g_{10}^2
\nonumber \\
&&
-~ \frac{331}{60 \Nc} g_1^2 g_9^2
- \frac{17}{3 \Nc} g_1^2 g_8^2
- \frac{4}{15} g_7^2 g_9^2
- \frac{56}{15} g_7^2 g_8^2
- \frac{2}{3} g_6^2 g_{11}^2
- \frac{8}{5} g_6^2 g_{10}^2
- \frac{24}{5} g_6^2 g_9^2
\nonumber \\
&&
-~ \frac{116}{3} g_6^2 g_8^2
- \frac{1}{15} g_5^2 g_{11}^2
- \frac{2}{15} g_5^2 g_{10}^2
- \frac{13}{15} g_5^2 g_9^2
- \frac{28}{3} g_5^2 g_8^2
- \frac{1}{30} g_4^2 g_{11}^2
- \frac{2}{15} g_4^2 g_{10}^2
\nonumber \\
&&
-~ \frac{1}{2} g_4^2 g_9^2
- \frac{82}{15} g_4^2 g_8^2
- \frac{31}{60} g_3^2 g_6^2
- \frac{3}{40} g_3^2 g_5^2
- \frac{1}{24} g_3^2 g_4^2
- \frac{13}{30} g_1 g_3 g_7^2
- \frac{43}{12} g_1 g_3 g_6^2
\nonumber \\
&&
-~ \frac{9}{10} g_1 g_3 g_5^2
- \frac{13}{20} g_1 g_3 g_4^2
- \frac{1}{30} g_1 g_2 g_7^2
- \frac{47}{20} g_1 g_2 g_6^2
- \frac{17}{30} g_1 g_2 g_5^2
- \frac{1}{3} g_1 g_2 g_4^2
+ \frac{17}{6} g_1^2 g_7^2
\nonumber \\
&&
+~ \frac{39}{2} g_1^2 g_6^2
+ \frac{617}{120} g_1^2 g_5^2
+ \frac{331}{120} g_1^2 g_4^2
- \frac{1}{4} \Nc g_{11}^4
- \frac{83}{30} \Nc g_{10}^2 g_{11}^2
- \frac{57}{10} \Nc g_{10}^4
- \frac{43}{60} \Nc g_9^2 g_{11}^2
\nonumber \\
&&
-~ \frac{229}{60} \Nc g_9^2 g_{10}^2
- \frac{41}{80} \Nc g_9^4
- \frac{27}{5} \Nc g_8^2 g_{11}^2
- \frac{106}{5} \Nc g_8^2 g_{10}^2
- \frac{223}{30} \Nc g_8^2 g_9^2
- \frac{302}{15} \Nc g_8^4
\nonumber \\
&&
-~ \frac{3}{160} \Nc g_3^2 g_{11}^2
- \frac{31}{240} \Nc g_3^2 g_{10}^2
- \frac{9}{320} \Nc g_3^2 g_9^2
- \frac{49}{240} \Nc g_3^2 g_8^2
- \frac{1}{7680} \Nc g_3^4
+ \frac{9}{40} \Nc g_2 g_3 g_8^2
\nonumber \\
&&
+~ \frac{27}{80} \Nc g_2^2 g_8^2
- \frac{61}{240} \Nc g_1 g_3 g_{11}^2
- \frac{223}{240} \Nc g_1 g_3 g_{10}^2
- \frac{169}{480} \Nc g_1 g_3 g_9^2
- \frac{11}{5} \Nc g_1 g_3 g_8^2
\nonumber \\
&&
-~ \frac{13}{1536} \Nc g_1 g_3^3
- \frac{13}{80} \Nc g_1 g_2 g_{11}^2
- \frac{31}{48} \Nc g_1 g_2 g_{10}^2
- \frac{107}{480} \Nc g_1 g_2 g_9^2
- \frac{211}{60} \Nc g_1 g_2 g_8^2
\nonumber \\
&&
+~ \frac{7}{3840} \Nc g_1 g_2 g_3^2
- \frac{7}{7680} \Nc g_1 g_2^2 g_3
+ \frac{143}{96} \Nc g_1^2 g_{11}^2
+ \frac{77}{15} \Nc g_1^2 g_{10}^2
+ \frac{1949}{960} \Nc g_1^2 g_9^2
\nonumber \\
&&
+~ \frac{19267}{1680} \Nc g_1^2 g_8^2
+ \frac{13}{2560} \Nc g_1^2 g_3^2
- \frac{259}{1920} \Nc g_1^2 g_2 g_3
- \frac{383}{7680} \Nc g_1^2 g_2^2
+ \frac{3961}{7680} \Nc g_1^3 g_3
\nonumber \\
&&
+~ \frac{889}{1280} \Nc g_1^3 g_2
- \frac{29269}{16128} \Nc g_1^4
+ \frac{1}{35} \Nf g_1^2 g_8^2
+ \frac{1}{10080} \Nf g_1^4 ~+~ O(g_i^6) \nonumber \\
\beta_9(g_i) &=&
\frac{70}{3 \Nc} g_{11}^4
+ \frac{116}{5 \Nc} g_{10}^2 g_{11}^2
+ \frac{56}{15 \Nc} g_{10}^4
+ \frac{34 \Nc} g_9^2 g_{11}^2
+ \frac{92}{5 \Nc} g_9^2 g_{10}^2
+ \frac{58}{3 \Nc} g_9^4
+ \frac{232}{15 \Nc} g_8^2 g_{11}^2
\nonumber \\
&&
+~ \frac{112}{15 \Nc} g_8^2 g_{10}^2
+ \frac{368}{15 \Nc} g_8^2 g_9^2
+ \frac{32}{5 \Nc} g_8^4
- \frac{1}{60 \Nc} g_3^2 g_{11}^2
+ \frac{1}{30 \Nc} g_3^2 g_{10}^2
+ \frac{1}{2 \Nc} g_3^2 g_9^2
\nonumber \\
&&
+~ \frac{2}{3 \Nc} g_3^2 g_8^2
+ \frac{1}{30 \Nc} g_1 g_3 g_{11}^2
+ \frac{22}{15 \Nc} g_1 g_3 g_{10}^2
- \frac{17}{30 \Nc} g_1 g_3 g_9^2
+ \frac{34}{15 \Nc} g_1 g_3 g_8^2
\nonumber \\
&&
+~ \frac{5}{6 \Nc} g_1 g_2 g_{11}^2
+ \frac{4}{15 \Nc} g_1 g_2 g_{10}^2
- \frac{3}{5 \Nc} g_1 g_2 g_9^2
- \frac{2}{5 \Nc} g_1 g_2 g_8^2
- \frac{41}{60 \Nc} g_1^2 g_{11}^2
\nonumber \\
&&
+~ \frac{31}{30 \Nc} g_1^2 g_{10}^2
+ \frac{13}{5 \Nc} g_1^2 g_9^2
+ \frac{48}{5 \Nc} g_1^2 g_8^2
- \frac{8}{15} g_7^2 g_{11}^2
- \frac{16}{15} g_7^2 g_{10}^2
- \frac{52}{15} g_7^2 g_9^2
- \frac{32}{15} g_7^2 g_8^2
\nonumber \\
&&
-~ \frac{8}{5} g_6^2 g_{11}^2
- \frac{4}{5} g_6^2 g_{10}^2
- \frac{16}{15} g_6^2 g_9^2
- \frac{8}{15} g_6^2 g_8^2
- \frac{26}{5} g_5^2 g_{11}^2
- \frac{46}{15} g_5^2 g_{10}^2
- 3 g_5^2 g_9^2
- \frac{4}{15} g_5^2 g_8^2
\nonumber \\
&&
-~ \frac{16}{15} g_4^2 g_{11}^2
- \frac{6}{5} g_4^2 g_{10}^2
- \frac{16}{5} g_4^2 g_9^2
- \frac{28}{15} g_4^2 g_8^2
- \frac{1}{3} g_3^2 g_7^2
- \frac{1}{60} g_3^2 g_6^2
+ \frac{1}{120} g_3^2 g_5^2
- \frac{1}{4} g_3^2 g_4^2
\nonumber \\
&&
-~ \frac{17}{15} g_1 g_3 g_7^2
- \frac{11}{15} g_1 g_3 g_6^2
- \frac{1}{60} g_1 g_3 g_5^2
+ \frac{17}{60} g_1 g_3 g_4^2
+ \frac{1}{5} g_1 g_2 g_7^2
- \frac{2}{15} g_1 g_2 g_6^2
\nonumber \\
&&
-~ \frac{5}{12} g_1 g_2 g_5^2
+ \frac{3}{10} g_1 g_2 g_4^2
- \frac{24}{5} g_1^2 g_7^2
- \frac{31}{60} g_1^2 g_6^2
+ \frac{41}{120} g_1^2 g_5^2
- \frac{13}{10} g_1^2 g_4^2
- \frac{23}{15} \Nc g_{11}^4
\nonumber \\
&&
-~ \frac{26}{15} \Nc g_{10}^2 g_{11}^2
- \frac{1}{3} \Nc g_{10}^4
- \frac{32}{15} \Nc g_9^2 g_{11}^2
- \frac{26}{15} \Nc g_9^2 g_{10}^2
- \frac{107}{60} \Nc g_9^4
- \frac{4}{15} \Nc g_8^2 g_{11}^2
\nonumber \\
&&
-~ \frac{4}{15} \Nc g_8^2 g_{10}^2
- \frac{8}{5} \Nc g_8^2 g_9^2
- \frac{4}{15} \Nc g_8^4
- \frac{1}{30} \Nc g_3^2 g_{11}^2
- \frac{3}{40} \Nc g_3^2 g_{10}^2
- \frac{11}{60} \Nc g_3^2 g_9^2
\nonumber \\
&&
-~ \frac{3}{20} \Nc g_3^2 g_8^2
- \frac{7}{1280} \Nc g_3^4
+ \frac{9}{40} \Nc g_2 g_3 g_9^2
+ \frac{27}{80} \Nc g_2^2 g_9^2
- \frac{1}{20} \Nc g_1 g_3 g_{11}^2
\nonumber \\
&&
-~ \frac{1}{6} \Nc g_1 g_3 g_{10}^2
- \frac{29}{120} \Nc g_1 g_3 g_9^2
+ \frac{13}{1920} \Nc g_1 g_3^3
-~ \frac{1}{5} \Nc g_1 g_2 g_{11}^2
- \frac{1}{30} \Nc g_1 g_2 g_{10}^2
\nonumber \\
&&
-~ \frac{34}{15} \Nc g_1 g_2 g_9^2
+ \frac{2}{15} \Nc g_1 g_2 g_8^2
+ \frac{1}{240} \Nc g_1 g_2 g_3^2
- \frac{1}{1920} \Nc g_1 g_2^2 g_3
- \frac{7}{60} \Nc g_1^2 g_{11}^2
\nonumber \\
&&
-~ \frac{47}{120} \Nc g_1^2 g_{10}^2
- \frac{583}{840} \Nc g_1^2 g_9^2
- \frac{91}{60} \Nc g_1^2 g_8^2
+ \frac{3}{320} \Nc g_1^2 g_3^2
+ \frac{59}{320} \Nc g_1^2 g_2 g_3
\nonumber \\
&&
+~ \frac{331}{1920} \Nc g_1^2 g_2^2
- \frac{161}{384} \Nc g_1^3 g_3
- \frac{331}{320} \Nc g_1^3 g_2
+~ \frac{87677}{80640} \Nc g_1^4
+ \frac{1}{35} \Nf g_1^2 g_9^2
+ \frac{5}{252} \Nf g_1^4 \nonumber \\
&&
+~ O(g_i^6) \nonumber \\
\beta_{10}(g_i) &=&
\frac{18}{5 \Nc} g_{11}^4
+ \frac{196}{5 \Nc} g_{10}^2 g_{11}^2
+ \frac{248}{3 \Nc} g_{10}^4
+ \frac{62}{15 \Nc} g_9^2 g_{11}^2
+ \frac{116}{5 \Nc} g_9^2 g_{10}^2
+ \frac{6}{5 \Nc} g_9^4
+ \frac{8}{3 \Nc} g_8^2 g_{11}^2
\nonumber \\
&&
+~ \frac{272}{15 \Nc} g_8^2 g_{10}^2
+ \frac{32}{15 \Nc} g_8^2 g_9^2
+ \frac{32}{15 \Nc} g_8^4
- \frac{3}{20 \Nc} g_3^2 g_{11}^2
- \frac{31}{30 \Nc} g_3^2 g_{10}^2
- \frac{1}{12 \Nc} g_3^2 g_9^2
\nonumber \\
&&
-~ \frac{9}{5 \Nc} g_1 g_3 g_{11}^2
- \frac{43}{6 \Nc} g_1 g_3 g_{10}^2
- \frac{13}{10 \Nc} g_1 g_3 g_9^2
- \frac{13}{15 \Nc} g_1 g_3 g_8^2
- \frac{17}{15 \Nc} g_1 g_2 g_{11}^2
\nonumber \\
&&
-~ \frac{47}{10 \Nc} g_1 g_2 g_{10}^2
- \frac{2}{3 \Nc} g_1 g_2 g_9^2
- \frac{1}{15 \Nc} g_1 g_2 g_8^2
+ \frac{617}{60 \Nc} g_1^2 g_{11}^2
+ \frac{39}{\Nc} g_1^2 g_{10}^2
+ \frac{331}{60 \Nc} g_1^2 g_9^2
\nonumber \\
&&
+~ \frac{17}{3 \Nc} g_1^2 g_8^2
- \frac{8}{15} g_7^2 g_{11}^2
- \frac{56}{15} g_7^2 g_{10}^2
- \frac{4}{15} g_7^2 g_9^2
- \frac{48}{5} g_6^2 g_{11}^2
- \frac{604}{15} g_6^2 g_{10}^2
- \frac{82}{15} g_6^2 g_9^2
\nonumber \\
&&
-~ \frac{16}{5} g_6^2 g_8^2
- \frac{26}{15} g_5^2 g_{11}^2
- \frac{142}{15} g_5^2 g_{10}^2
- \frac{14}{15} g_5^2 g_9^2
- \frac{4}{15} g_5^2 g_8^2
- g_4^2 g_{11}^2
- \frac{28}{5} g_4^2 g_{10}^2
\nonumber \\
&&
-~ \frac{8}{15} g_4^2 g_9^2
- \frac{4}{15} g_4^2 g_8^2
+ \frac{31}{60} g_3^2 g_6^2
+ \frac{3}{40} g_3^2 g_5^2
+ \frac{1}{24} g_3^2 g_4^2
+ \frac{13}{30} g_1 g_3 g_7^2
+ \frac{43}{12} g_1 g_3 g_6^2
\nonumber \\
&&
+~ \frac{9}{10} g_1 g_3 g_5^2
+ \frac{13}{20} g_1 g_3 g_4^2
+ \frac{1}{30} g_1 g_2 g_7^2
+ \frac{47}{20} g_1 g_2 g_6^2
+ \frac{17}{30} g_1 g_2 g_5^2
+ \frac{1}{3} g_1 g_2 g_4^2
- \frac{17}{6} g_1^2 g_7^2
\nonumber \\
&&
-~ \frac{39}{2} g_1^2 g_6^2
- \frac{617}{120} g_1^2 g_5^2
- \frac{331}{120} g_1^2 g_4^2
- \frac{1}{30} \Nc g_{11}^4
- \frac{1}{5} \Nc g_{10}^2 g_{11}^2
- \frac{4}{15} \Nc g_{10}^4
- \frac{1}{30} \Nc g_9^2 g_{11}^2
\nonumber \\
&&
-~ \frac{1}{10} \Nc g_9^2 g_{10}^2
- \frac{1}{120} \Nc g_9^4
- \frac{2}{15} \Nc g_8^2 g_{11}^2
- \frac{8}{15} \Nc g_8^2 g_{10}^2
- \frac{1}{15} \Nc g_8^2 g_9^2
+ \frac{1}{240} \Nc g_3^2 g_{11}^2
\nonumber \\
&&
+~ \frac{13}{240} \Nc g_3^2 g_{10}^2
+ \frac{1}{480} \Nc g_3^2 g_9^2
+ \frac{9}{40} \Nc g_2 g_3 g_{10}^2
+ \frac{27}{80} \Nc g_2^2 g_{10}^2
+ \frac{17}{120} \Nc g_1 g_3 g_{11}^2
\nonumber \\
&&
-~ \frac{37}{120} \Nc g_1 g_3 g_{10}^2
+ \frac{17}{240} \Nc g_1 g_3 g_9^2
+ \frac{1}{10} \Nc g_1 g_3 g_8^2
- \frac{29}{7680} \Nc g_1 g_3^3
+ \frac{1}{120} \Nc g_1 g_2 g_{11}^2
\nonumber \\
&&
-~ \frac{289}{120} \Nc g_1 g_2 g_{10}^2
+ \frac{1}{240} \Nc g_1 g_2 g_9^2
- \frac{1}{15} \Nc g_1 g_2 g_8^2
- \frac{13}{1280} \Nc g_1 g_2 g_3^2
+ \frac{7}{7680} \Nc g_1 g_2^2 g_3
\nonumber \\
&&
+~ \frac{53}{240} \Nc g_1^2 g_{11}^2
+ \frac{1879}{1680} \Nc g_1^2 g_{10}^2
+ \frac{53}{480} \Nc g_1^2 g_9^2
- \frac{3}{5} \Nc g_1^2 g_8^2
- \frac{47}{7680} \Nc g_1^2 g_3^2
\nonumber \\
&&
+~ \frac{71}{960} \Nc g_1^2 g_2 g_3
+ \frac{259}{7680} \Nc g_1^2 g_2^2
- \frac{499}{7680} \Nc g_1^3 g_3
- \frac{1661}{3840} \Nc g_1^3 g_2
+ \frac{112391}{161280} \Nc g_1^4
\nonumber \\
&&
+~ \frac{1}{35} \Nf g_1^2 g_{10}^2
+ \frac{17}{5040} \Nf g_1^4 ~+~ O(g_i^6) \nonumber \\
\beta_{11}(g_i) &=&
\frac{14}{15 \Nc} g_{11}^4
+ \frac{28}{15 \Nc} g_{10}^2 g_{11}^2
+ \frac{8}{3 \Nc} g_{10}^4
+ \frac{54}{5 \Nc} g_9^2 g_{11}^2
+ \frac{68}{15 \Nc} g_9^2 g_{10}^2
+ \frac{6}{5 \Nc} g_9^4
+ \frac{40}{3 \Nc} g_8^2 g_{11}^2
\nonumber \\
&&
+~ \frac{112}{15 \Nc} g_8^2 g_{10}^2
+ \frac{32}{15 \Nc} g_8^2 g_9^2
+ \frac{32}{15 \Nc} g_8^4
+ \frac{1}{60 \Nc} g_3^2 g_{11}^2
- \frac{1}{30 \Nc} g_3^2 g_{10}^2
- \frac{1}{2 \Nc} g_3^2 g_9^2
\nonumber \\
&&
-~ \frac{2}{3 \Nc} g_3^2 g_8^2
- \frac{1}{30 \Nc} g_1 g_3 g_{11}^2
- \frac{22}{15 \Nc} g_1 g_3 g_{10}^2
+ \frac{17}{30 \Nc} g_1 g_3 g_9^2
- \frac{34}{15 \Nc} g_1 g_3 g_8^2
\nonumber \\
&&
-~ \frac{5}{6 \Nc} g_1 g_2 g_{11}^2
- \frac{4}{15 \Nc} g_1 g_2 g_{10}^2
+ \frac{3}{5 \Nc} g_1 g_2 g_9^2
+ \frac{2}{5 \Nc} g_1 g_2 g_8^2
+ \frac{41}{60 \Nc} g_1^2 g_{11}^2
\nonumber \\
&&
-~ \frac{31}{30 \Nc} g_1^2 g_{10}^2
- \frac{13}{5 \Nc} g_1^2 g_9^2
- \frac{48}{5 \Nc} g_1^2 g_8^2
- \frac{16}{5} g_7^2 g_{11}^2
- \frac{16}{15} g_7^2 g_{10}^2
- \frac{4}{15} g_7^2 g_9^2
- \frac{4}{15} g_6^2 g_{11}^2
\nonumber \\
&&
-~ \frac{4}{15} g_6^2 g_{10}^2
- \frac{4}{5} g_6^2 g_9^2
- \frac{8}{15} g_6^2 g_8^2
- \frac{2}{5} g_5^2 g_{11}^2
- \frac{2}{15} g_5^2 g_{10}^2
- \frac{13}{5} g_5^2 g_9^2
- \frac{44}{15} g_5^2 g_8^2
\nonumber \\
&&
-~ \frac{8}{3} g_4^2 g_{11}^2
- \frac{14}{15} g_4^2 g_{10}^2
- \frac{8}{15} g_4^2 g_9^2
- \frac{4}{15} g_4^2 g_8^2
+ \frac{1}{3} g_3^2 g_7^2
+ \frac{1}{60} g_3^2 g_6^2
- \frac{1}{120} g_3^2 g_5^2
\nonumber \\
&&
+~ \frac{1}{4} g_3^2 g_4^2
+ \frac{17}{15} g_1 g_3 g_7^2
+ \frac{11}{15} g_1 g_3 g_6^2
+ \frac{1}{60} g_1 g_3 g_5^2
- \frac{17}{60} g_1 g_3 g_4^2
- \frac{1}{5} g_1 g_2 g_7^2
\nonumber \\
&&
+~ \frac{2}{15} g_1 g_2 g_6^2
+ \frac{5}{12} g_1 g_2 g_5^2
- \frac{3}{10} g_1 g_2 g_4^2
+ \frac{24}{5} g_1^2 g_7^2
+ \frac{31}{60} g_1^2 g_6^2
- \frac{41}{120} g_1^2 g_5^2
\nonumber \\
&&
+~ \frac{13}{10} g_1^2 g_4^2
+ \frac{7}{15} \Nc g_{11}^4
+ \frac{2}{15} \Nc g_{10}^2 g_{11}^2
- \frac{4}{15} \Nc g_{10}^4
- \frac{8}{15} \Nc g_9^2 g_{11}^2
- \frac{4}{15} \Nc g_9^2 g_{10}^2
\nonumber \\
&&
+~ \frac{7}{60} \Nc g_9^4
- \frac{4}{5} \Nc g_8^2 g_{11}^2
- \frac{8}{15} \Nc g_8^2 g_{10}^2
+ \frac{4}{15} \Nc g_8^2 g_9^2
- \frac{1}{16} \Nc g_3^2 g_{11}^2
- \frac{1}{40} \Nc g_3^2 g_{10}^2
\nonumber \\
&&
+~ \frac{13}{240} \Nc g_3^2 g_9^2
+ \frac{1}{12} \Nc g_3^2 g_8^2
+ \frac{1}{384} \Nc g_3^4
+ \frac{9}{40} \Nc g_2 g_3 g_{11}^2
+ \frac{27}{80} \Nc g_2^2 g_{11}^2
- \frac{3}{8} \Nc g_1 g_3 g_{11}^2
\nonumber \\
&&
+~ \frac{7}{15} \Nc g_1 g_3 g_{10}^2
- \frac{1}{12} \Nc g_1 g_3 g_9^2
+ \frac{1}{5} \Nc g_1 g_3 g_8^2
+ \frac{1}{384} \Nc g_1 g_3^3
- \frac{53}{24} \Nc g_1 g_2 g_{11}^2
\nonumber \\
&&
+~ \frac{1}{12} \Nc g_1 g_2 g_{10}^2
- \frac{1}{8} \Nc g_1 g_2 g_9^2
- \frac{1}{10} \Nc g_1 g_2 g_8^2
+ \frac{1}{1920} \Nc g_1 g_2^2 g_3
- \frac{361}{1680} \Nc g_1^2 g_{11}^2
\nonumber \\
&&
+~ \frac{23}{120} \Nc g_1^2 g_{10}^2
+ \frac{71}{240} \Nc g_1^2 g_9^2
+ \frac{83}{60} \Nc g_1^2 g_8^2
+ \frac{13}{1920} \Nc g_1^2 g_3^2
- \frac{63}{320} \Nc g_1^2 g_2 g_3
\nonumber \\
&&
-~ \frac{111}{640} \Nc g_1^2 g_2^2
+ \frac{631}{1920} \Nc g_1^3 g_3
+ \frac{1001}{960} \Nc g_1^3 g_2
- \frac{23629}{20160} \Nc g_1^4
+ \frac{1}{35} \Nf g_1^2 g_{11}^2
\nonumber \\
&&
-~ \frac{149}{5040} \Nf g_1^4 ~+~ O(g_i^6) ~.
\end{eqnarray}
Again these renormalization group functions, as well as those for $SU(3)$,
satisfy the same checks we discussed for the $SU(2)$ case.  

\sect{Large $\Nf$ check.}

We devote this section to the final independent check we have on the 
renormalization group functions in each of the three cases which is the 
comparison with the large $\Nf$ critical exponents which have been computed in
the non-abelian Thirring model universality class. The background to this is 
the observation that the renormalization group functions depend on the
parameter $\Nf$ and the various coupling constants for a specific value of
$\Nc$. The coefficients of these parameters in each renormalization group
function is conventionally determined by perturbative methods as was carried
out in the previous section. However one can also determine the coefficients
via an ordering of graphs defined by $\Nf$. This is achieved through the known 
$d$-dependent critical exponents of the underlying universality class. An 
alternative view of this is that the exponents already contain information on 
the perturbative coefficients. The method is to compute the renormalization 
group functions at the Wilson-Fisher fixed point in 
$d$~$=$~$8$~$-$~$2\epsilon$, expand in powers of $1/\Nf$ and then compare with 
the $\epsilon$ expansion of the corresponding large $\Nf$ critical exponents. 
This constitutes our independent check. The first step in the procedure is to 
locate the Wilson-Fisher fixed point explicitly order by order in powers of 
$1/\Nf$ and $\epsilon$ by finding the solution to
\begin{equation}
\beta_i(g_j) ~=~ 0
\label{betazero}
\end{equation}
for the $d$-dimensional $\beta$-functions. In four dimensions this is 
relatively straightforward since there is only one coupling constant in QCD. 
For eight dimensions we have $11$ coupling constants for the case of $SU(\Nc)$.
So we follow the method introduced in \cite{22,23}. As there are $3$- and 
$4$-leg operators in (\ref{lagqcd8}) we have to be careful in defining the 
rescaling which is the initial step in the approach of \cite{22,23}. Therefore 
at the outset we set
\begin{eqnarray}
g_i &=& \sqrt{\frac{70 \epsilon}{\Nf}} x_i ~~~~~~ \mbox{$i$ ~=~ 1 to 3}
\nonumber \\
g_i^2 &=& \frac{70 \epsilon}{\Nf} x_i ~~~~~~~~~ \mbox{$i$ ~=~ 4 to 11} ~.
\label{critcoup}
\end{eqnarray}
in (\ref{betazero}) and expand in powers of $\epsilon$ and $1/\Nf$. First the
leading order term in $1/\Nf$ of the equations is isolated and then the 
$\epsilon$ expansion of this leading term is found before repeating the 
exercise for the subsequent term in the large $\Nf$ expansion. For 
the $SU(\Nc)$ $\beta$-functions the resulting critical couplings are
\begin{eqnarray}
x_1 &=& 1 ~+~ \frac{1933 \Nc}{24 \Nf} ~+~ \frac{3736489 \Nc^2}{384 \Nf^2} ~+~ 
O \left( \epsilon ; \frac{1}{\Nf^3} \right) \nonumber \\ 
x_2 &=& \frac{17}{9} ~+~ \frac{287279 \Nc}{1944 \Nf} ~+~ 
\frac{5066611513 \Nc^2}{279936 \Nf^2} ~+~ 
O \left( \epsilon ; \frac{1}{\Nf^3} \right) \nonumber \\ 
x_3 &=& \frac{16}{3} ~+~ \frac{143411 \Nc}{324 \Nf} ~+~ 
\frac{153781987 \Nc^2}{2916 \Nf^2} ~+~ 
O \left( \epsilon ; \frac{1}{\Nf^3} \right) \nonumber \\
x_4 &=& -~ \frac{25}{9 \Nc} ~-~ 
\left[ \frac{1615081}{46656} + \frac{115591}{432 \Nc^2} \right] 
\frac{1}{\Nf} \nonumber \\
&& +~ \left[ \frac{4084305085}{11664 \Nc^3}
- \frac{318375286621}{839808 \Nc} + \frac{894758019623 \Nc}{3359232} \right] 
\frac{1}{\Nf^2} ~+~ O \left( \epsilon ; \frac{1}{\Nf^3} \right) \nonumber \\
x_5 &=& \frac{149}{36 \Nc} ~+~ \left[ \frac{39472453}{46656}
- \frac{343}{432 \Nc^2} \right] \frac{1}{\Nf} \nonumber \\
&& +~ 
\left[ - \frac{768922651}{11664 \Nc^3} + \frac{6468807373}{839808 \Nc}
+ \frac{144625900963 \Nc}{1119744} \right] \frac{1}{\Nf^2} ~+~ 
O \left( \epsilon ; \frac{1}{\Nf^3} \right) \nonumber \\ 
x_6 &=& \frac{149}{72} ~+~ \left[ \frac{18279803 \Nc}{46656}
- \frac{343}{432 \Nc} \right] \frac{1}{\Nf} \nonumber \\
&& +~ \left[ 
\frac{161927831371 \Nc^2}{2239488} - \frac{28392695975}{1679616}
- \frac{1571677793}{46656 \Nc^2} \right] \frac{1}{\Nf^2} ~+~ 
O \left( \epsilon ; \frac{1}{\Nf^3} \right) \nonumber \\ 
x_7 &=& -~ \frac{17}{36 \Nc} ~-~ \left[ \frac{997943}{46656}
+ \frac{343}{432 \Nc^2} \right] \frac{1}{\Nf} \nonumber \\
&& +~ \left[ 
\frac{285596549}{2916 \Nc^3} + \frac{155421633577}{839808 \Nc}
+ \frac{69408246905 \Nc}{1119744} \right] \frac{1}{\Nf^2} ~+~ 
O \left( \epsilon ; \frac{1}{\Nf^3} \right) \nonumber \\ 
x_8 &=& -~ \frac{1}{72 \Nc} ~+~ \left[ \frac{343}{216 \Nc^2} 
- \frac{5517727}{46656} \right] \frac{1}{\Nf} \nonumber \\
&& +~ \left[
\frac{37377424567}{209952 \Nc} - \frac{20021939}{23328 \Nc^3}
- \frac{223918424851 \Nc}{3359232} \right] \frac{1}{\Nf^2} ~+~ 
O \left( \epsilon ; \frac{1}{\Nf^3} \right) \nonumber \\ 
x_9 &=& -~ \frac{1}{144} ~+~ \left[ \frac{343}{216 \Nc} 
- \frac{6535889 \Nc}{186624} \right] \frac{1}{\Nf} \nonumber \\
&& +~ \left[
\frac{388161667565}{3359232} - \frac{5239709503}{93312 \Nc^2}
- \frac{297237914233 \Nc^2}{26873856} \right] \frac{1}{\Nf^2} ~+~ 
O \left( \epsilon ; \frac{1}{\Nf^3} \right) \nonumber \\ 
x_{10} &=& -~ \frac{17}{72} ~-~ \left[ \frac{343}{432 \Nc}
+ \frac{2137045 \Nc}{46656} \right] \frac{1}{\Nf} \nonumber \\
&& -~ \left[
\frac{11234911345}{104976} + \frac{2318604883}{46656 \Nc^2}
+ \frac{7914271411 \Nc^2}{497664} \right] \frac{1}{\Nf^2} ~+~ 
O \left( \epsilon ; \frac{1}{\Nf^3} \right) \nonumber \\ 
x_{11} &=& -~ \frac{25}{18} ~-~ \left[ \frac{115591}{432 \Nc}
+ \frac{8299843 \Nc}{93312} \right] \frac{1}{\Nf} \nonumber \\
&& +~ 
\left[ \frac{15724650809}{46656 \Nc^2} - \frac{429314818345}{1679616}
+ \frac{261182511995 \Nc^2}{6718464} \right] \frac{1}{\Nf^2} ~+~ 
O \left( \epsilon ; \frac{1}{\Nf^3} \right)
\label{critvals}
\end{eqnarray}
where the double order symbol indicates both the two loop correction and the
next order in the large $\Nf$ expansion. These values of $x_i$ correspond to 
the $\epsilon$ expansion of all the critical couplings to the order which they 
are known in the previous section. Next the renormalization group functions for
the wave function renormalization are evaluated at the Wilson-Fisher critical 
point and expanded in powers of both $\epsilon$ and $1/\Nf$. Subsequently the 
critical exponents should be in agreement with the coefficients of $\epsilon$ 
in the known large $\Nf$ critical exponents of the non-abelian Thirring 
universality class when they are expanded around $d$~$=$~$8$~$-$~$2\epsilon$. 
Substituting the values from (\ref{critvals}) into (\ref{rgesunc}) we find for
$SU(\Nc)$ that 
\begin{eqnarray}
\left. \gamma_A(g_c) \right|_{\alpha=0} &=& \epsilon ~+~ 
\frac{245 \Nc}{12 \Nf} \epsilon ~+~ \frac{473585 \Nc^2}{144 \Nf^2} \epsilon ~+~
O \left( \epsilon^2 ; \frac{1}{\Nf^3} \right) \nonumber \\
\left. \gamma_c(g_c) \right|_{\alpha=0} &=& 
-~ \frac{245 \Nc}{24 \Nf} \epsilon ~-~ 
\frac{473585 \Nc^2}{288 \Nf^2} \epsilon ~+~ 
O \left( \epsilon^2 ; \frac{1}{\Nf^3} \right) \nonumber \\
\left. \gamma_\psi(g_c) \right|_{\alpha=0} &=&
\left[ \frac{245 \Nc}{12} - \frac{245}{12 \Nc} \right] \frac{\epsilon}{\Nf} ~+~
\left[ \frac{473585 \Nc^2}{144} - \frac{473585}{144} \right] 
\frac{\epsilon}{\Nf^2} ~+~ O \left( \epsilon^2 ; \frac{1}{\Nf^3} \right)
\label{critdim}
\end{eqnarray}
where $g_c$ denotes the set of critical couplings defined in (\ref{critcoup}).
In order to compare with the large $\Nf$ critical exponents of the universal 
theory founded on the non-abelian Thirring model at the Wilson-Fisher fixed 
point we have to restrict the exponents to the Landau gauge. This is because in
effect the gauge parameter $\alpha$ acts as an additional coupling constant and
the Landau gauge is the corresponding fixed point in this context. In other 
words the gauge dependent large $\Nf$ critical exponents of the gluon, quark 
and ghost fields can only be compared with the Landau gauge anomalous 
dimensions at criticality which has been noted before in \cite{16,18}. We 
restrict our large $\Nf$ comparison to these three anomalous dimensions since 
they are the only three quantities which are available for eight dimensional 
QCD. While the large $\Nf$ critical exponent of the four dimensional QCD 
$\beta$-function is known at $O(1/\Nf)$, \cite{16}, that exponent would relate 
to the renormalization of the operator $\frac{1}{4} G_{\mu\nu}^a 
G^{a \, \mu\nu}$ in (\ref{lagqcd8m}). In four dimensions the gauge coupling 
constant in four dimensional QCD is dimensionless but in the continuation along
the thread of the $d$-dimensional Wilson-Fisher fixed point the coupling 
becomes dimensionful and the correction to scaling exponent in four dimensions 
transcends into a mass parameter in higher dimensions such as the eight 
dimensional Lagrangian (\ref{lagqcd8m}). Therefore, if we evaluate the leading 
order $d$-dimensional large $\Nf$ critical exponents for the gluon, quark and 
ghost fields of \cite{51} near eight dimensions by setting 
$d$~$=$~$8$~$-$~$2\epsilon$ we find that the coefficients of $\epsilon$ match 
precisely with those of (\ref{critdim}) in the Landau gauge for $SU(\Nc)$. 
Moreover, since the quark anomalous dimension is also known at $O(1/\Nf^2)$ in 
the Landau gauge, \cite{18}, it is satisfying to record that the corresponding 
term of $\left. \gamma_\psi(g_c) \right|_{\alpha=0}$ is in full agreement. 
While we have not given explicit details for the $SU(2)$ and $SU(3)$ 
renormalization group functions we note that we have carried out the same check
as $SU(\Nc)$ and found that there is full consistency in these cases too. 
Consequently the ultraviolet completion of QCD or the non-abelian Thirring 
model to eight dimensions via (\ref{lagqcd8}) has been established at one loop 
within the large $\Nf$ expansion as expected.  

\sect{Dimension $8$ operators in four dimensions.}

In this section we turn to a complementary problem which is the renormalization
of dimension $8$ operators in four dimensions. Such operators in the case of
Yang-Mills theory have been considered in \cite{30,32} where, for instance, the 
anomalous dimensions for the $SU(2)$ and $SU(3)$ groups were computed at one 
loop in \cite{32}. The reason for this is that in four dimensions the canonical 
dimensions of the gluon and ghost fields are such that there is a complicated 
mixing between gluonic and quark operators. In (\ref{lagqcd8}) by contrast on
dimensional grounds it is not possible to have any other interactions involving
quarks aside from the quark-gluon interaction. Therefore in this section we
concentrate on the renormalization of four dimensional dimension $8$ operators
in $SU(\Nc)$ Yang-Mills theory for $\Nc$~$\geq$~$4$ as this case has not been
considered. In addition we use the {\em same} operator basis as was used in 
(\ref{lagqcd8}), which differs from that of \cite{30,32}, in order to ease
structural comparisons. First, to set notation the basis for the dimension $8$ 
operators in four dimensions for the colour group $SU(\Nc)$ we use is
\begin{eqnarray}
{\cal O}_{841} &=& G_{\mu \sigma}^a G^{a \, \mu \rho} G^{b \, \sigma \nu} 
G_{\rho \nu}^b ~~~,~~~ 
{\cal O}_{842} ~=~ G_{\mu \sigma}^a G^{b \, \mu \rho} G^{b \, \sigma \nu} 
G_{\rho \nu}^a \nonumber \\
{\cal O}_{843} &=& G_{\mu \sigma}^a G_{\nu \rho}^a G^{b \, \sigma \mu} 
G^{b \, \rho \nu} ~~~,~~~
{\cal O}_{844} ~=~ G_{\mu \sigma}^a G_{\nu \rho}^b G^{a \, \sigma \mu} 
G^{b \, \rho \nu} \nonumber \\
{\cal O}_{845} &=& d_4^{a b c d} G_{\mu \sigma}^a G^{b \, \mu \sigma} 
G_{\nu \rho}^c G^{d \, \nu \rho} ~~~,~~~
{\cal O}_{846} ~=~ d_4^{a b c d} G_{\mu \sigma}^a G^{c \, \mu \rho} 
G^{b \, \nu \sigma} G_{\nu \rho}^d \nonumber \\
{\cal O}_{847} &=& d_4^{a c b d} G_{\mu \sigma}^a G^{b \, \mu \sigma} 
G_{\nu \rho}^c G^{d \, \nu \rho} ~~~,~~~ 
{\cal O}_{848} ~=~ d_4^{a d b c} G_{\mu \sigma}^a G^{c \, \mu \rho} 
G^{b \, \nu \sigma} G_{\nu \rho}^d ~.
\label{dim8op}
\end{eqnarray}
The notation is similar to that used in \cite{32}. However, these operators are
{\em not} the same since we have specified the basis with respect to a specific
colour group unlike \cite{32}. We have chosen this ordering so that the $SU(2)$
basis corresponds to the first four operators and that for $SU(3)$ involves the
first six. Equally the ordering is equivalent to that used in (\ref{lagqcd8}) 
for the quartic gluon interactions with coupling constants $g_4$ to $g_{11}$
respectively. 

To renormalize the operators ${\cal O}_{84i}$ we use the same technique as that
for the $4$-point functions of (\ref{lagqcd8}) but in this case we apply it to
the Green's function $\langle A_\mu^a(p_1) A_\nu^b(p_2) A_\sigma^c(p_3) 
A_\rho^d(p_4) {\cal O}_{84i}(p_5) \rangle$ where 
$p_5$~$=$~$-$~$\sum_{i=1}^4 p_i$. However, as we are considering
an operator renormalization there will be a mixing of the ${\cal O}_{84i}$
operators among themselves which will produce a mixing matrix of anomalous
dimensions. This is similar to the $\beta$-functions for the couplings in
(\ref{lagqcd8}). However for operator renormalization there are aspects to
address compared with a Lagrangian renormalization. For instance, for the gauge
invariant dimension $8$ operators (\ref{dim8op}) there will be mixing into 
gauge variant and equation of motion operators as well as possibly total 
derivative operators. The latter can arise when an operator is renormalized in 
a Green's function where the insertion is at non-zero momentum insertion. 
Moreover this set includes total derivative operators which are gauge 
invariant, gauge variant and equation of motion operators. So the mixing matrix
in effect is larger than an  $8$~$\times$~$8$ matrix based on (\ref{dim8op}). 
Not only do the operators of (\ref{dim8op}) mix with all operators of the 
enlarged set but the gauge variant, equation of motion and total derivative 
operators can mix with themselves when each is renormalized. However, the 
overall mixing matrix has a particular structure in that the gauge invariant 
operators mix with all classes of operators but the gauge variant ones only mix
within that class. See, for instance, \cite{52,53,54,55}. As we are primarily 
interested in the gauge invariant operators we restrict the evaluation of the 
Green's function $\langle A_\mu^a(p_1) A_\nu^b(p_2) A_\sigma^c(p_3) 
A_\rho^d(p_4) {\cal O}_{84i}(p_5) \rangle$ to the case where the external gluon
legs are all on-shell. The condition for a gluon $A^a_\mu(p)$ to be on-shell is
that its polarization vector and momentum satisfy 
\begin{equation}
p_\mu p^\mu ~=~ 0 ~~~~,~~~~ p^\mu \epsilon_\mu(p) ~=~ 0 ~.
\label{onshell}
\end{equation}
Therefore we multiply the Green's function by $\epsilon^\mu(p_1) 
\epsilon^\nu(p_2) \epsilon^\sigma(p_3) \epsilon^\rho(p_4)$ and apply 
(\ref{onshell}). The terms which remain such as $\epsilon_\mu(p_i) p_j^\mu$ for
$i$~$\neq$~$j$ or $p_i p_j$ are resolved by grouping them in terms 
corresponding to the Feynman rules of the contributing operators such as 
(\ref{dim8op}) and any gauge invariant total derivative or equation of motion 
operators. The reason why this list omits gauge variant operators is that the 
restriction of (\ref{onshell}) corresponds to taking a physical matrix element.
As such no gauge variant operators can be present, \cite{52,53,54,55}.

Necessary to achieve the resolution into this basis of operators is that the 
operator has to be inserted at non-zero momentum. If it was inserted at zero 
momentum then certain terms of the Feynman rule of different operators will be 
similar and hence the extraction of the renormalization constants in the mixing
matrix cannot be achieved uniquely and unambiguously. Therefore, formally the 
set of bare operators, denoted by the subscript ${}_{\mbox{\footnotesize{o}}}$ 
satisfy
\begin{equation}
{\cal O}_{i \, \mbox{\footnotesize{o}}} ~=~ Z_{ij} {\cal O}_j
\end{equation}
where $Z_{ij}$ is the mixing matrix of renormalization constants from which the
mixing matrix of anomalous dimensions, $\gamma_{ij}(a)$, can be deduced. In
this section $a$~$=$~$g^2/(16\pi^2)$ denotes the coupling constant of four 
dimensional QCD where $g$ is the coupling present in the covariant derivative. 
It transpires that for the eight operators (\ref{dim8op}) the matrix needs to 
be enlarged since there is mixing into an equation of motion operator. In 
\cite{32} the seven independent equation of motion operators were constructed 
and are
\begin{eqnarray}
{\cal O}_{82e1} &=& 
D^\mu G^a_{\mu\nu} D^\rho D_\sigma D_\rho G^{a\,\nu\sigma} ~~~,~~~ 
{\cal O}_{82e2} ~=~ 
D^\sigma D^\mu G^a_{\mu\nu} D^\rho D^\nu G^a_{\sigma\rho} \nonumber \\
{\cal O}_{82e3} &=& 
D^\sigma D^\mu G^a_{\mu\nu} D_\rho D_\sigma G^{a\,\nu\rho} ~~~,~~~
{\cal O}_{82e4} ~=~ 
D_\sigma G^a_{\nu\rho} D^\sigma D^\rho D_\mu G^{a\,\mu\nu} \nonumber \\
{\cal O}_{82e5} &=& 
G^a_{\nu\sigma} D^\sigma D^\rho D_\rho D_\mu G^{a\,\mu\nu} \nonumber \\
{\cal O}_{83e1} &=& f^{abc} 
G^a_{\sigma\rho} D^\nu G^{b\,\sigma\rho} D^\mu G^c_{\mu\nu} ~~~,~~~
{\cal O}_{83e2} ~=~ f^{abc} 
G^{a~\nu}_{~\sigma} G^{b\,\sigma\rho} D_\rho D^\mu G^c_{\mu\nu}
\label{dim8eomop}
\end{eqnarray}
where the first two labels indicate the operator dimension and gluon leg number
respectively and note that each operator is gauge invariant. We recall that in 
four dimensions the equation of motion of the gluon in Yang-Mills theory is
\begin{equation}
D^\mu G_{\mu\nu} ~=~ 0 
\end{equation}
which is relatively simple in contrast to that of (\ref{lagqcd8}). Unlike 
(\ref{dim8op}) there is no reduction of the equation of motion set
(\ref{dim8eomop}) depending on which colour group we consider. One comment is 
in order with respect to (\ref{lagqcd8}) which is that the operators 
(\ref{dim8eomop}) are not present in that Lagrangian. The reason why they are 
considered part of the basis here arises from the different nature of the two
types of renormalizations we are carrying out. In (\ref{lagqcd8}) for the purely
gluonic sector we included the set of independent gauge invariant operators 
involving the field strength. The operators which were dependent, and hence not
included, were equivalent to linear combinations of the ones appearing in 
(\ref{lagqcd8}) as well as operators which were total derivatives. In a 
Lagrangian context the latter operators can be integrated out and hence were 
not included in (\ref{lagqcd8}). For the renormalization of the dimension $8$ 
operators (\ref{dim8op}) in four dimensions one has to accommodate mixing into 
the various operator classes noted earlier. As one of these classes involves 
equation of motion operators we have included these in the set of operators for
our mixing. However it is a straightforward exercise to show that the operators
${\cal O}_{82ei}$ can each be related to the gluon kinetic operator plus higher
leg operators and those with a total derivative. Equally the operators 
${\cal O}_{83ei}$ in eight dimensions can be mapped to the operators with 
couplings $g_2$ and $g_3$ respectively plus higher leg and total derivative 
operators in (\ref{lagqcd8}).

The final stage of the operator renormalization is the evaluation of the 
divergent part of the on-shell Green's function. Like the renormalization of
the $4$-point functions of (\ref{lagqcd8}) we apply the vacuum bubble
expansion based on (\ref{vacbubexp}). The only major difference between its
use here and the previous application is that after the expansion and the
Laporta reduction the master integral is evaluated in {\em four} dimensions. 
Therefore, extracting the renormalization constants we find the elements of the
mixing matrix are 
\begin{eqnarray}
\gamma_{841,841}(a) &=& \frac{8}{3\Nc} a \,+\, O(a^2) ~,~
\gamma_{841,842}(a) ~=~ -~ \frac{8}{3\Nc} a \,+\, O(a^2) \nonumber \\
\gamma_{841,843}(a) &=& \frac{22}{3\Nc} a \,+\, O(a^2) ~,~
\gamma_{841,844}(a) ~=~ -~ \frac{1}{6\Nc} [ 11 \Nc^2 + 44 ] a \,+\, O(a^2) 
\nonumber \\
\gamma_{841,845}(a) &=& -~ \frac{11}{3} a \,+\, O(a^2) ~,~
\gamma_{841,846}(a) ~=~ \frac{4}{3} a \,+\, O(a^2) \nonumber \\
\gamma_{841,847}(a) &=& \frac{11}{3} a \,+\, O(a^2) ~,~
\gamma_{841,848}(a) ~=~ -~ \frac{4}{3} a \,+\, O(a^2) \nonumber \\
\gamma_{842,841}(a) &=& -~ \frac{1}{3\Nc} [ 14 \Nc^2 + 4 ] a \,+\, O(a^2) ~,~
\gamma_{842,842}(a) ~=~ -~ \frac{1}{3\Nc} [ 10 \Nc^2 - 4 ] a \,+\, O(a^2) 
\nonumber \\
\gamma_{842,843}(a) &=& \frac{1}{3\Nc} [ 12 \Nc^2 + 22 ] a \,+\, O(a^2) ~,~
\gamma_{842,844}(a) ~=~ -~ \frac{1}{6\Nc} [ - \Nc^2 + 44 ] a \,+\, O(a^2) 
\nonumber \\
\gamma_{842,845}(a) &=& -~ \frac{11}{3} a \,+\, O(a^2) ~,~
\gamma_{842,846}(a) ~=~ -~ \frac{2}{3} a \,+\, O(a^2) \nonumber \\
\gamma_{842,847}(a) &=& \frac{11}{3} a \,+\, O(a^2) ~,~
\gamma_{842,848}(a) ~=~ \frac{2}{3} a \,+\, O(a^2) \nonumber \\
\gamma_{843,841}(a) &=& -~ \frac{1}{3\Nc} [ 28 \Nc^2 + 68 ] a \,+\, O(a^2) ~,~
\gamma_{843,842}(a) ~=~ -~ \frac{1}{3\Nc} [ - 24 \Nc^2 - 68 ] a \,+\, O(a^2) 
\nonumber \\
\gamma_{843,843}(a) &=& \frac{1}{3\Nc} [ 2 \Nc^2 + 50 ] a \,+\, O(a^2) ~,~
\gamma_{843,844}(a) ~=~ -~ \frac{1}{3\Nc} [ - \Nc^2 + 50 ] a \,+\, O(a^2) 
\nonumber \\
\gamma_{843,845}(a) &=& -~ \frac{25}{3} a \,+\, O(a^2) ~,~
\gamma_{843,846}(a) ~=~ -~ \frac{34}{3} a \,+\, O(a^2) \nonumber \\
\gamma_{843,847}(a) &=& \frac{25}{3} a \,+\, O(a^2) ~,~
\gamma_{843,848}(a) ~=~ \frac{34}{3} a \,+\, O(a^2) \nonumber \\
\gamma_{844,841}(a) &=& -~ \frac{56}{\Nc} a \,+\, O(a^2) ~,~
\gamma_{844,842}(a) ~=~ \frac{56}{\Nc} a \,+\, O(a^2) \nonumber \\
\gamma_{844,843}(a) &=& -~ \frac{4}{\Nc} a \,+\, O(a^2) ~,~
\gamma_{844,844}(a) ~=~ -~ \frac{1}{3\Nc} [ 22 \Nc^2 - 12 ] a \,+\, O(a^2) 
\nonumber \\
\gamma_{844,845}(a) &=& 2 a \,+\, O(a^2) ~,~
\gamma_{844,846}(a) ~=~ -~ 28 a \,+\, O(a^2) \nonumber \\
\gamma_{844,847}(a) &=& -~ 2 a \,+\, O(a^2) ~,~
\gamma_{844,848}(a) ~=~ 28 a \,+\, O(a^2) \nonumber \\
\gamma_{845,841}(a) &=& -~ \frac{1}{\Nc^2} [ 28 \Nc^2 - 112 ] a \,+\, 
O(a^2) ~,~
\gamma_{845,842}(a) ~=~ \frac{1}{\Nc^2} [ 28 \Nc^2 - 112 ] a \,+\, O(a^2) 
\nonumber \\
\gamma_{845,843}(a) &=& -~ \frac{1}{\Nc^2} [ 2 \Nc^2 - 8 ] a \,+\, O(a^2) ~,~
\gamma_{845,844}(a) ~=~ -~ \frac{1}{\Nc^2} [ - 2 \Nc^2 + 8 ] a \,+\, O(a^2) 
\nonumber \\
\gamma_{845,845}(a) &=& -~ \frac{1}{2\Nc} [ 5 \Nc^2 + 8 ] a \,+\, O(a^2) ~,~
\gamma_{845,846}(a) ~=~ -~ \frac{1}{\Nc} [ 6 \Nc^2 - 56 ] a \,+\, O(a^2) 
\nonumber \\
\gamma_{845,847}(a) &=& -~ \frac{1}{3\Nc} [ 2 \Nc^2 - 12 ] a \,+\, O(a^2) ~,~
\gamma_{845,848}(a) ~=~ -~ \frac{1}{3\Nc} [ - 16 \Nc^2 + 168 ] a \,+\, O(a^2) 
\nonumber \\
\gamma_{846,841}(a) &=& -~ \frac{1}{3\Nc^2} [ - 4 \Nc^2 + 16 ] a \,+\, 
O(a^2) ~,~
\gamma_{846,842}(a) ~=~ -~ \frac{1}{3\Nc^2} [ 4 \Nc^2 - 16 ] a \,+\, O(a^2) 
\nonumber \\
\gamma_{846,843}(a) &=& -~ \frac{1}{3\Nc^2} [ - 11 \Nc^2 + 44 ] a \,+\, 
O(a^2) ~,~
\gamma_{846,844}(a) ~=~ -\, \frac{1}{3\Nc^2} [ 11 \Nc^2 - 44 ] a \,+\, O(a^2) 
\nonumber \\
\gamma_{846,845}(a) &=& -~ \frac{1}{3\Nc} [ 4 \Nc^2 - 22 ] a \,+\, O(a^2) ~,~
\gamma_{846,846}(a) ~=~ -~ \frac{1}{3\Nc} [ 3 \Nc^2 + 8 ] a \,+\, O(a^2) 
\nonumber \\
\gamma_{846,847}(a) &=& -~ \frac{1}{3\Nc} [ - 3 \Nc^2 + 22 ] a \,+\, 
O(a^2) ~,~
\gamma_{846,848}(a) ~=~ -~ \frac{1}{3\Nc} [ 3 \Nc^2 - 8 ] a \,+\, O(a^2) 
\nonumber \\
\gamma_{847,841}(a) &=& -~ \frac{1}{3\Nc^2} [ 34 \Nc^2 - 136 ] a \,+\, O(a^2) 
\nonumber \\
\gamma_{847,842}(a) &=& -~ \frac{1}{3\Nc^2} [ - 34 \Nc^2 + 136 ] a \,+\, O(a^2) 
\nonumber \\
\gamma_{847,843}(a) &=& -~ \frac{1}{3\Nc^2} [ - 25 \Nc^2 + 100 ] a \,+\, O(a^2) 
\nonumber \\
\gamma_{847,844}(a) &=& -~ \frac{1}{3\Nc^2} [ 25 \Nc^2 - 100 ] a \,+\, O(a^2) 
\nonumber \\
\gamma_{847,845}(a) &=& -~ \frac{1}{12\Nc} [ 25 \Nc^2 - 200 ] a \,+\, 
O(a^2) ~,~
\gamma_{847,846}(a) ~=~ -~ \frac{1}{3\Nc} [ 19 \Nc^2 - 68 ] a \,+\, O(a^2) 
\nonumber \\
\gamma_{847,847}(a) &=& -~ \frac{1}{3\Nc} [ - 3 \Nc^2 + 50 ] a \,+\, O(a^2) ~,~ 
\gamma_{847,848}(a) ~=~ -~ \frac{1}{3\Nc} [ - 16 \Nc^2 + 68 ] a \,+\, O(a^2) 
\nonumber \\
\gamma_{848,841}(a) &=& -~ \frac{1}{3\Nc^2} [ 2 \Nc^2 - 8 ] a \,+\, O(a^2) ~,~
\gamma_{848,842}(a) ~=~ -~ \frac{1}{3\Nc^2} [ - 2 \Nc^2 + 8 ] a \,+\, O(a^2) 
\nonumber \\
\gamma_{848,843}(a) &=& -~ \frac{1}{3\Nc^2} [ - 11 \Nc^2 + 44 ] a \,+\, O(a^2) 
\nonumber \\
\gamma_{848,844}(a) &=& -~ \frac{1}{3\Nc^2} [ 11 \Nc^2 - 44 ] a \,+\, O(a^2) 
\nonumber \\
\gamma_{848,845}(a) &=& -~ \frac{1}{6\Nc} [ 5 \Nc^2 - 44 ] a \,+\, O(a^2) ~,~
\gamma_{848,846}(a) ~=~ -~ \frac{1}{3\Nc} [ 8 \Nc^2 - 4 ] a \,+\, O(a^2) 
\nonumber \\
\gamma_{848,847}(a) &=& -~ \frac{1}{3\Nc} [ - 6 \Nc^2 + 22 ] a \,+\, O(a^2) 
\nonumber \\
\gamma_{848,848}(a) &=& -~ \frac{1}{3\Nc} [ 4 \Nc^2 + 4 ] a \,+\, O(a^2) 
\end{eqnarray}
for $SU(\Nc)$. For the eight $SU(\Nc)$ dimension $8$ core operators at one loop
there is mixing into only one equation of motion operator which is 
${\cal O}_{83e2}$. More explicitly we have 
\begin{eqnarray}
\gamma_{841,83e2}(a) &=& -~ 2 a \,+\, O(a^2) ~,~
\gamma_{842,83e2}(a) ~=~ 4 a \,+\, O(a^2) ~,~
\gamma_{843,83e2}(a) ~=~ 4 a \,+\, O(a^2) \nonumber \\ 
\gamma_{844,83e2}(a) &=& -~ 8 a \,+\, O(a^2) ~,~
\gamma_{845,83e2}(a) ~=~ -~ \frac{4}{\Nc} [ \Nc^2 - 4 ] a \,+\, O(a^2) 
\nonumber \\
\gamma_{846,83e2}(a) &=& -~ \frac{1}{\Nc} [ \Nc^2 - 4 ] a \,+\, O(a^2) ~,~
\gamma_{847,83e2}(a) ~=~ \frac{2}{\Nc} [ \Nc^2 - 4 ] a \,+\, O(a^2) 
\nonumber \\
\gamma_{848,83e2}(a) &=& \frac{2}{\Nc} [ \Nc^2 - 4 ] a \,+\, O(a^2) ~.
\label{mixmateom}
\end{eqnarray}
The mixing of the main operators into this specific equation of motion operator
is necessary as otherwise divergences would remain in each of the Green's 
functions. In other words there are not sufficient counterterms and freedom
available from the set of operators in (\ref{dim8op}) alone to obtain a finite
expression. For $SU(2)$ and $SU(3)$ the respective parts for this sector of the
mixing matrix are contained within (\ref{mixmateom}). For $SU(2)$ only the 
first four operators of (\ref{dim8op}) are active and for $SU(3)$ it is the
first six. Then for $SU(2)$ the first four entries in (\ref{mixmateom}) 
correspond to the $4$-leg operator mixing into the equation of motion 
operators. Clearly $\gamma_{845,83e2}(a)$ vanishes for $\Nc$~$=$~$2$ as a 
consistency check. The situation for $SU(3)$ is similar except the first six 
entries are relevant but $\Nc$~$=$~$3$ has to be set. Finally, the equation of 
motion operators can mix with themselves and we have determined that sector of 
the mixing matrix in the same way by inserting each operator in the physical 
matrix element. The only non-zero entries are 
\begin{equation}
\gamma_{83e1,82e4}(a) ~=~ -~ \frac{1}{3\Nc} a \,+\, O(a^2) ~~~,~~~
\gamma_{83e1,82e5}(a) ~=~ \frac{1}{2\Nc} a \,+\, O(a^2) 
\end{equation}
which is valid for all the $SU(\Nc)$ groups. This completes our dimension $8$ 
operator analysis in four dimensions for the particular $SU(\Nc)$ colour 
groups. These results together with the $SU(2)$ and $SU(3)$ cases are all 
included in the data file. While this is a fully separate computation to the 
renormalization of (\ref{lagqcd8}) the structural parallels of the respective 
renormalization group functions are now evident. 

\sect{Discussion.}

One of our main goals was to construct the eight dimensional quantum field
theory which was in the same universality class as the two dimensional 
non-abelian Thirring model and four dimensional QCD at their respective 
Wilson-Fisher fixed points. We have managed to achieve this by following the
guiding principles established for the parallel construction for scalar field
theories with an $O(N)$ symmetry. The first of these is to retain the core 
interaction between the matter and force fields which in the present case were
a spin-$\half$ fermion and spin-$1$ boson field in the adjoint representation
of the colour group. This interaction is the only one present in the base
theory of the tower of theories lying in the universality class which is the
non-abelian Thirring model, \cite{21}. The second aspect is renormalizability. 
This means that extra interactions have to be included in the critical 
dimension of each of the subsequent Lagrangians of the tower so that each 
Lagrangian is renormalizable. These extra independent operators, which are 
purely gluonic for this universality class, will become irrelevant or relevant 
away from the critical dimension. So for example including the canonical gluon
kinetic operator for QCD in the non-abelian Thirring model would render it 
nonrenormalizable in two dimensions. The final main principle is the 
requirement of gauge fixing. We chose a linear covariant gauge fixing in order 
to make connections with lower dimensional results and extended the 
Faddeev-Popov construction to eight dimensions. This last step is necessary as 
the two dimensional non-abelian Thirring model has a conserved current, 
$\bar{\psi} \gamma^\mu T^a \psi$, whose $2$-point correlation function is 
transverse. While there is no gluon as such in the non-abelian Thirring model, 
like the four dimensional gauge theory case, the field $A^a_\mu$ is an 
auxiliary in two dimensions and corresponds to this current. In other words the
correlation of $A^a_\mu$ in two dimensions is in effect akin to a Landau gauge 
propagator. As the gauge parameter, $\alpha$, in QCD is effectively a second 
coupling constant then at criticality one has to effect its critical coupling 
which corresponds in fact to the Landau gauge. This accords with the 
establishment of (\ref{lagqcd8}) as being in same universality class as the 
non-abelian Thirring model and QCD via the large $\Nf$ expansion. One can only 
compare the $d$-dimensional large $\Nf$ critical exponents with the exponents 
derived from gauge dependent renormalization group functions when the 
$\epsilon$ expansion of the latter have been computed in the Landau gauge. We 
have checked this off explicitly here for eight dimensional QCD from the one 
loop renormalization group functions. Put another way the Wilson-Fisher fixed 
point underlying this particular universality class preserves the 
transversality of the gluon across the dimensions. 

There are several future avenues to pursue in light of our analysis. One is to 
build the ten dimensional theory of a spin-$1$ field coupled to a fundamental
fermion which lies in the non-abelian Thirring model universality class. The 
procedure to do this evidently follows the above outline. It would have no 
technical obstacles aside from the calculational one of requiring a large 
amount of integration by parts to determine even just the one loop 
renormalization group functions. This will be a tedious exercise rather than an
insurmountable problem. Another obvious extension is to construct the 
renormalization group functions of (\ref{lagqcd8}) at two loops. Indeed this 
has already been achieved for QED, \cite{27,26}. However in eight dimensions 
the computations were manageable due to there being only four independent 
interactions and more crucially no quintic or sextic gauge interactions. These 
were obviously present in the non-abelian case and also increased the amount of 
integration needed in order to evaluate the large number of Feynman graphs with
high exponent gluon propagators, \cite{26}. With the tower of Lagrangians 
essentially established at the Wilson-Fisher fixed point for the non-abelian 
Thirring model universality class the next focus ought to be on the connection 
of non-Lagrangian operators in the universal theory. These operators will have 
massive couplings in the non-critical dimensions but are relevant in 
constructing effective field Lagrangians in a specific dimension. In other 
words there should be a drive to study the operator anomalous dimensions at 
criticality. 

We have taken the first step in this direction by renormalizing dimension $8$ 
operators in four dimensions. While laying the foundation to this here by 
illustrating the structural parallels of the renormalization group functions, 
the next step is to introduce quark contributions. These are required for the 
large $\Nf$ expansion connection where the underlying operator critical
exponents in the universal theory would also need to be found in addition to 
the mixing matrices in perturbation theory. The perturbative computations to
construct such mixing matrices should not be regarded as a straightforward 
task. One reason for this is due to the canonical dimensions of the quark and 
gluon fields being different in $d$-dimensions. Hence quark and gluon operators
will have different canonical dimensions except in one particular dimension. 
Therefore we did not have to consider what would ordinarily be dimension $8$
quark operators in the four dimensional sense in the construction of the eight 
dimensional Lagrangian (\ref{lagqcd8}). However, in four dimensional QCD there 
are dimension $8$ operators with quark content in addition to the gluon 
operators of (\ref{dim8op}). This was one of the reasons why our focus was on 
Yang-Mills operators here as an exploratory exercise in the context of 
(\ref{lagqcd8}) and to observe that the structure of the respective four and 
eight dimensional renormalization group functions were not dissimilar. While 
(\ref{lagqcd8}) has a quark operator it is the kinetic term and it does not 
have the same canonical dimension as, say, the operators of (\ref{dim8op}) in 
four dimensions. The first stage in such an investigation will be to set up the 
large $\Nf$ formalism for dimension $6$ and $8$ gauge invariant operators and 
compute the mixing matrix of critical exponents at $O(1/\Nf)$ in 
$d$-dimensions. The former dimension is required for an analysis of 
(\ref{lagqcd6}) and we note that the large $\Nf$ exponent relating to the QCD 
$\beta$-function in four dimensions, \cite{16}, was derived from the critical
point large $\Nf$ renormalization of the dimension four operator $G^a_{\mu\nu} 
G^{a\,\mu\nu}$. That in effect was the initial step of the proposal to examine 
the operator content of the tower of Lagrangians constituting universal 
non-abelian Thirring model universality class.

\vspace{1cm}
\noindent
{\bf Acknowledgements.} This work was carried out with the support of the STFC 
through the Consolidated Grant ST/L000431/1. The author thanks K.L. Jones, R.M.
Simms and Dr. Marco Bochicchio for useful discussions.


\begin{thebibliography}{99}
\bibitem{1} D.J. Gross \& F.J. Wilczek, Phys. Rev. Lett. {\bf 30} (1973), 1343.
\bibitem{2} H.D. Politzer, Phys. Rev. Lett. {\bf 30} (1973), 1346.
\bibitem{3} W.E. Caswell, Phys. Rev. Lett. {\bf 33} (1974), 244.
\bibitem{4} D.R.T. Jones, Nucl. Phys. {\bf B75} (1974), 531.
\bibitem{5} O.V. Tarasov, A.A. Vladimirov \& A.Yu. Zharkov, Phys. Lett.
{\bf B93} (1980), 429.
\bibitem{6} T. van Ritbergen, J.A.M. Vermaseren \& S.A. Larin, Phys. Lett.
{\bf B400} (1997), 379.
\bibitem{7} M. Czakon, Nucl. Phys. {\bf B710} (2005), 485.
\bibitem{8} P.A. Baikov, K.G. Chetyrkin \& J.H. K\"{u}hn, JHEP {\bf 1410}
(2014), 76.
\bibitem{9} P.A. Baikov, K.G. Chetyrkin \& J.H. K\"{u}hn, Phys. Rev. Lett.
{\bf 118} (2017), 082002.
\bibitem{10} F. Herzog, B. Ruijl, T. Ueda, J.A.M. Vermaseren \& A. Vogt, JHEP
{\bf 1702} (2017), 090.
\bibitem{11} T. Luthe, A. Maier, P. Marquard \& Y. Schr\"{o}der, JHEP 
{\bf 1701} (2017), 081.
\bibitem{12} T. Luthe, A. Maier, P. Marquard \& Y. Schr\"{o}der, JHEP 
{\bf 1703} (2017), 020.
\bibitem{13} P.A. Baikov, K.G. Chetyrkin \& J.H. K\"{u}hn, JHEP {\bf 1704}
(2017), 119.
\bibitem{14} T. Luthe, A. Maier, P. Marquard \& Y. Schr\"{o}der, JHEP 
{\bf 1710} (2017), 166.
\bibitem{15} K.G. Chetyrkin, G. Falcioni, F. Herzog \& J.A.M. Vermaseren, JHEP 
{\bf 1710} (2017), 179.
\bibitem{16} J.A. Gracey, Phys. Lett. {\bf B373} (1996), 178.
\bibitem{17} A. Palanques-Mestre \& P. Pascual, Commun. Math. Phys. {\bf 95}
(1984), 277.
\bibitem{18} M. Ciuchini, S.\'{E}. Derkachov, J.A. Gracey \& A.N. Manashov, 
Nucl. Phys. {\bf B579} (2000), 56.
\bibitem{19} A.N. Vasil'ev, Y.M. Pismak \& J.R. Honkonen, Theor. Math. Phys. 
{\bf 46} (1981), 104.
\bibitem{20} A.N. Vasil'ev, Y.M. Pismak \& J.R. Honkonen, Theor. Math. Phys. 
{\bf 47} (1981), 465.
\bibitem{21} A. Hasenfratz \& P. Hasenfratz, Phys. Lett. {\bf B297} (1992),
166.
\bibitem{22} L. Fei, S. Giombi \& I.R. Klebanov, Phys. Rev. {\bf D90} (2014),
025018.
\bibitem{23} L. Fei, S. Giombi, I.R. Klebanov \& G. Tarnopolsky, Phys. Rev.
{\bf D91} (2015), 045011.
\bibitem{24} J.A. Gracey, Phys. Rev. {\bf D92} (2015), 025012.
\bibitem{25} J.A. Gracey \& R.M. Simms, Phys. Rev. {\bf D96} (2017), 025022. 
\bibitem{26} J.A. Gracey, Phys. Rev. {\bf D93} (2016), 025025.
\bibitem{27} S. Giombi, I.R. Klebanov \& G. Tarnopolsky, J. Phys. {\bf A49}
(2016), 134503. 
\bibitem{28} A.J. McKane, D.J. Wallace \& R.K.P. Zia, Phys. Lett. {\bf B65}
(1976), 171.
\bibitem{29} A.J. McKane, J. Phys. {\bf G3} (1977), 1165.
\bibitem{30} A.Yu. Morozov, Sov. J. Nucl. Phys. {\bf 40} (1984), 505.
\bibitem{31} D.I. Kazakov, JHEP {\bf 0303} (2002), 020.
\bibitem{32} J.A. Gracey, Nucl. Phys. {\bf B634} (2002), 192; Nucl. Phys.
{\bf B696} (2004), 295. 
\bibitem{33} A.J. Macfarlane, A. Sudbery \& P.H. Weisz, Commun. Math. Phys.
{\bf 11} (1968), 77.
\bibitem{34} M. Stingl, Phys. Rev. {\bf D34} (1986), 3863; Phys. Rev. {\bf D36}
(1987), 651(E).
\bibitem{35} P. Nogueira, J. Comput. Phys. {\bf 105} (1993), 279.
\bibitem{36} S.A. Larin \& J.A.M. Vermaseren, Phys. Lett. {\bf B303} (1993),
334.
\bibitem{37} J.A.M. Vermaseren, math-ph/0010025.
\bibitem{38} M. Tentyukov \& J.A.M. Vermaseren, Comput. Phys. Commun. {\bf 181}
(2010), 1419.
\bibitem{39} S. Laporta, Int. J. Mod. Phys. {\bf A15} (2000), 5087.
\bibitem{40} C. Studerus, Comput. Phys. Commun. {\bf 181} (2010), 1293.
\bibitem{41} A. von Manteuffel \& C. Studerus, arXiv:1201.4330. 
\bibitem{42} O.V. Tarasov, Phys. Rev. {\bf D54} (1996), 6479.
\bibitem{43} O.V. Tarasov, Nucl. Phys. {\bf B502} (1997), 455.
\bibitem{44} P.A. Baikov \& K.G. Chetyrkin, Nucl. Phys. {\bf B837} (2010), 186.
\bibitem{45} N.I. Ussyukina \& A.I. Davydychev, Phys. Atom. Nucl. {\bf 56}
(1993), 1553.
\bibitem{46} N.I. Ussyukina \& A.I. Davydychev, Phys. Lett. {\bf B332} (1994),
159.
\bibitem{47} N.I. Ussyukina \& A.I. Davydychev, Phys. Lett. {\bf B305} (1993),
136.
\bibitem{48} M. Misiak \& M. M\"{u}nz, Phys. Lett. {\bf B344} (1995), 308.
\bibitem{49} K.G. Chetyrkin, M. Misiak \& M. M\"{u}nz, Nucl. Phys. {\bf B518}
(1998), 473.
\bibitem{50} S. Weinberg, Phys. Rev. {\bf 118} (1960), 838.
\bibitem{51} J.A. Gracey, Phys. Lett. {\bf B318} (1993), 177.
\bibitem{52} S.D. Joglekar \& B.W. Lee, Ann. Phys. {\bf 97} (1976), 160.
\bibitem{53} S.D. Joglekar, Ann. Phys. {\bf 100} (1976), 395.
\bibitem{54} S.D. Joglekar, Ann. Phys. {\bf 108} (1977), 233.
\bibitem{55} S.D. Joglekar, Ann. Phys. {\bf 109} (1977), 210.
\end{thebibliography}
\end{document}